\documentclass[3p,times,twocolumn]{elsarticle}

\usepackage{ecrc-mb}
\usepackage{widetext}

\volume{00}

\firstpage{1}

\journalname{Nuclear Physics B Proceedings Supplement}

\runauth{M. Beneke}
\runtitle{Soft-collinear factorization in $B$ decays}
\runprep{TUM-HEP 978/15, SFB/CPP-14-119, 
18 January 2015}

\jid{nuphbp}

\jnltitlelogo{Nuclear Physics B Proceedings Supplement}

\usepackage{amssymb}

\usepackage[figuresright]{rotating}

\newcommand{\npslash}{{n \!\!\! /}_+ }
\newcommand{\nmslash}{{n \!\!\! /}_- }

\newcommand{\Wcp}{W_{c2}}
\newcommand{\gp}{\gamma_{\perp}}

\newcommand{\OII}{{O^{\mathrm{II}}}}
\newcommand{\nph}{{\frac{\npslash}{2}}}
\newcommand{\nmh}{{\frac{\nmslash}{2}}}

\newcommand{\xib}{{\bar \xi}}
\def\Slash#1{#1 \hskip-0.59em /}
\newcommand{\ov}[1]{ \overleftarrow{#1} }
\def\np{n_+}

\def\J{{\cal J}}

\begin{document}

\begin{frontmatter}



\dochead{}


\title{Soft-collinear factorization in $B$ decays}


\author{M. Beneke}

\address{Physik Department T31, 
James-Franck-Stra\ss e~1, 
Technische Universit\"at M\"unchen,
D--85748 Garching, Germany}

\begin{abstract}
\noindent
The combination of collinear factorization with effective field theory 
originally developed for soft interactions of heavy quarks provides the 
foundations of the theory of exclusive and semi-inclusive $B$ decays. 
In this article I summarize some of the later conceptual developments 
of the so-called QCD factorization approach that make use of 
soft-collinear effective theory. Then I discuss the status and results 
of the calculation of the hard-scattering functions at the next order, 
and review very briefly some of the phenomenology, covering aspects of 
charmless, electroweak 
penguin and radiative (semi-leptonic) decays. 
\end{abstract}

\begin{keyword}
Flavour physics \sep $B$-meson \sep charmless decays  \sep  
QCD factorization \sep CP violation\\[0.0cm]


\end{keyword}

\end{frontmatter}


\section{Introduction}
\label{sec:intro}

In the early 1990s the heavy-quark effective theory 
\cite{Eichten:1989zv,Grinstein:1990mj,Georgi:1990um} and the 
heavy-quark expansion \cite{Shifman:1986sm,Shifman:1987rj,Bigi:1993fe} 
were developed into powerful tools to describe 
the dynamics and emerging symmetries \cite{Isgur:1989vq,Isgur:1989ed} 
of QCD in the limit, when 
the mass $m_b$ of a quark is much larger than the intrinsic scale 
$\Lambda$ of the strong interaction. The basic assumption for 
the validity of this expansion is that the limit is taken when 
all other dimensionful parameters, including the momenta of 
particles, are fixed to be of order $\Lambda$ or another soft 
scale. After integrating out the scale $m_b$, the 
effective dynamics is governed exclusively by the soft scale, 
since no other scale is present in the problem.

In the second half of the 1990s the CLEO experiment at Cornell 
accumulated enough statistics of $B$ mesons to measure for 
the first time the rare $b\to u$ and loop-induced decays in 
exclusive two-body final states. The tools above do not apply to 
this situation. When the final state consists of two light mesons, 
say two pions, the energy of the pions is roughly $m_b/2$. 
The presence of a large external momentum invalidates the assumption 
that all small-virtuality fluctuations are soft. Energetic, light  
particles can radiate energetic particles and remain close to 
their mass-shell when the radiation is collinear to their direction 
of flight. When the invariant mass squared is of order $\Lambda^2$ as must 
be the case for an exclusive final state of light mesons, collinear 
radiation cannot be described perturbatively.

The idea of collinear factorization is central to the QCD treatment of 
high-energy scattering with classic applications to deep-inelastic 
scattering, exclusive form factors at high-momentum transfer, and 
jet physics. In these cases, a crucial step is often the demonstration 
that soft effects cancel, such that the non-perturbative, collinear 
physics can be isolated in parton distributions and light-cone 
distribution amplitudes. The then novel experimental accessibility of 
exclusive, energetic 
final states in $B$-meson decays required the extension of collinear 
factorization to situations where the large energy is injected into 
the process not by a colourless external source such as the electromagnetic 
current, but the weak decay of a heavy quark surrounded by the soft 
degrees of freedom that make up the $B$ meson. Soft physics is therefore 
expected to be more relevant than in traditional applications of 
collinear factorization.

\begin{figure}
\vskip0.2cm
\begin{center}
\includegraphics[width=7.5cm]{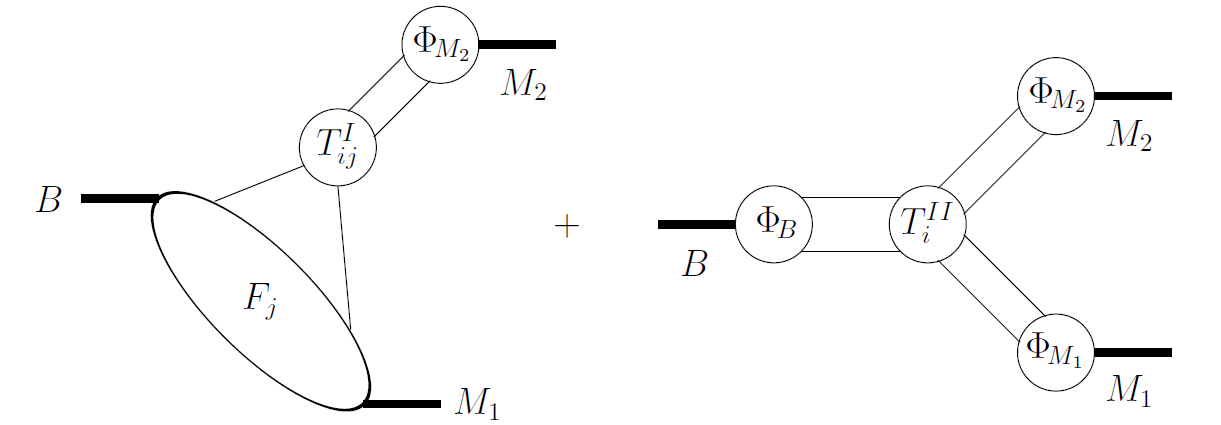}
\end{center}
\vspace*{-0.3cm}
\caption{
\label{fig:factformula}
Graphical representation of the factorization formula. Only one of 
the two form-factor terms in 
(\ref{fff}) is shown for simplicity. Figure from \cite{Beneke:2000ry}.}
\end{figure}

The problem was first addressed in a systematic way by Buchalla, Neubert, 
Sachrajda and myself~\cite{Beneke:1999br,Beneke:2000ry,Beneke:2001ev}, 
motivated by the search for a rigorous QCD 
treatment of exclusive heavy-light ($D\pi$) and light-light (
$\pi\pi$, ``charmless'') final states, based on nothing but the 
scale hierarchy $m_b\gg \Lambda$ and the concepts of perturbative 
soft and collinear factorization. Assuming that the electroweak 
scale $M_W$ is already integrated out, the starting point is the 
weak effective Lagrangian, which contains the flavour-changing 
interactions in the form of local dimension-six operators $Q_i$ of the 
four-fermion or chromo- and electromagnetic dipole type. 
The relevant objects to 
describe the $B\to M_1 M_2$ decay amplitude are then the matrix 
elements $\langle M_1 M_2|Q_i|\bar{B}\rangle$ of these 
operators. For the further discussion below, I quote the factorization 
formula, valid in the heavy-quark limit up to corrections of order 
$\Lambda/m_b$, in the form originally given 
in~\cite{Beneke:1999br,Beneke:2000ry}:
\begin{eqnarray}
\label{fff}
\langle M_1 M_2|Q_i|\bar{B}\rangle &=& 
\nonumber\\
&&\hspace*{-2.3cm}
\sum_j F_j^{B\to M_1}(m_2^2)\,\int_0^1 du\,T_{ij}^{\rm I}(u)\,\phi_{M_2}(u) 
\nonumber\\
&&\hspace*{-2.3cm}
+\,\,(M_1\leftrightarrow M_2)\nonumber\\
&&\hspace*{-2.3cm}
+\int_0^1 \!d\xi du dv \,T_i^{\rm II}(\omega,u,v)\,
\phi_B(\xi)\,\phi_{M_1}(v)\,\phi_{M_2}(u)  \nonumber\\
&&\hspace*{-1.5cm}\mbox{if $M_1$ and $M_2$ are both light,}\\[0.2cm]
\langle M_1 M_2|Q_i|\bar{B}\rangle &=& 
\nonumber\\
&&\hspace*{-2.3cm}
\sum_j F_j^{B\to M_1}(m_2^2)\,\int_0^1 du\,T_{ij}^{\rm I}(u)\,\phi_{M_2}(u) 
\label{fff2}
\nonumber\\
&&\hspace*{-1.5cm}\mbox{if $M_1$ is heavy and $M_2$ is light.} 
\end{eqnarray} 
Here $F_j^{B\to M_{1,2}}(m_{2,1}^2)$ denotes a $B\to M_{1,2}$ form factor, 
and $\phi_X(u)$ is the light-cone distribution amplitude (LCDA) for the 
quark-antiquark Fock state of meson $X$. 
$T_{ij}^{\rm I}(u)$ and $T_i^{\rm II}(\xi,u,v)$ are hard-scattering functions, 
which are calculable perturbatively in the strong coupling $\alpha_s$;  
$m_{1,2}$ denote the light meson masses.  
Eq.~(\ref{fff}) is represented graphically in 
Figure~\ref{fig:factformula}. The presence of a 
form factor and of LCDAs shows that soft {\em and} collinear physics 
is relevant even at the leading order in the heavy-quark expansion.

Eq.~(\ref{fff}) applies to decays into two 
light me\-sons, for which the spectator quark in the $B$ meson 
can go to either of the final-state mesons. An example is the decay 
$B^-\to \pi^0 K^-$. If the spectator quark can go only to one of the 
final-state mesons, as for example in $\bar B_d\to \pi^+ K^-$, 
we call this meson $M_1$ and the second form-factor term related to 
$M_1\leftrightarrow M_2$ 
on the right-hand side of (\ref{fff}) is absent. 
The factorization formula simplifies when the spectator quark goes 
to a heavy meson [see (\ref{fff2})], 
such as in $\bar{B}_d\to D^+ \pi^-$. In this case 
the hard interactions with the spectator quark represented by the 
third term in (\ref{fff}) can be dropped, because
they are power-suppressed in the heavy-quark limit. 
In the opposite situation that 
the spectator quark goes to a light meson but the other meson is heavy, 
factorization does not hold. 

The simplest case to gain some intuition is $\bar{B}_d\to D^+\pi^-$, 
when the $D$ meson is also assumed to be parametrically heavy. The 
spectator quark and other light degrees of freedom in the $B$ meson 
have to rearrange themselves only slightly to form a $D$ meson 
together with the charm quark created in the weak $b\to c \bar u d$ 
transition. For the other two light quarks $\bar u d$ to form a 
pion with energy $\mathcal{O}(m_b)$, they must be highly energetic and 
collinear, and in a colour-singlet configuration. Soft interactions 
decouple from such a configuration and this allows it to leave 
the decay region without interfering with the $D$ meson 
formation~\cite{Bjorken:1988kk}.  The probability of such a special 
configuration to form a pion is described by the leading-twist pion 
LCDA $\phi_\pi(u)$. The factorization formula (\ref{fff2})  
provides the quantitative framework for this discussion, which allows 
to compute higher-order corrections. 
For example, if the light quark-antiquark pair is initially formed 
in a colour-octet state, one can still show that soft gluons 
decouple at leading order in $\Lambda/m_b$, 
if this pair is to end up as a pion. This implies that 
the pair must interact with a {\em hard} gluon, and hence this provides a 
calculable strong-interaction correction to the basic mechanism 
discussed above.\footnote{A correction of this type was computed already 
in~\cite{Politzer:1991au}, but the generality and 
importance of the result went unnoticed.} An important element in 
demonstrating the suppression of soft interactions, except for 
those parametrized by the $B\to D$ form factor, is the assumption 
that the pion LCDA vanishes at least linearly as the longitudinal momentum 
fraction approaches the endpoints $u=0,1$. This assumption can be 
justified by the fact that it is satisfied 
by the asymptotic distribution amplitude $\phi_\pi(u)= 6 u (1-u)$, 
which is the appropriate one in the infinite heavy-quark mass limit. If this 
were not the case, the pion could be formed with larger probability 
in an asymmetric configuration, in which one constituent has  
soft momentum of order $\Lambda$. Soft gluons would not decouple from 
this soft constituent, thus spoiling factorization.

The above discussion relies crucially on the spectator 
quark in the $B$ meson 
going to the heavy meson in the final state. If, as in the 
case of a $D^0 \pi^0$ final state, the spectator quark must be 
picked up by the light meson, the amplitude is suppressed by the 
$B\to \pi$ form factor. But since the $D$ meson's size is of 
order $1/\Lambda$, the $D^0$ formation and $B\to \pi$ 
transition cannot be assumed to not interfere and factorization 
is violated. 

The case of two light final state mesons is the most interesting 
one. The dominant decay process is indeed the same as for the 
case of the $D^+ \pi^-$ final state, but this implies that the 
light meson that picks up the spectator quark is formed in a very 
asymmetric configuration in which the spectator quark carries a tiny 
fraction $\Lambda/m_b$ of the total momentum of the light meson. Such a 
configuration is suppressed by the endpoint behaviour of the LCDA 
as discussed above, but owing to this
suppression there exists a competing process, in which a hard 
gluon is exchanged with the spectator quark, propelling it to 
large energy, thus avoiding the endpoint-suppression penalty factor. 
If the hard gluon connects to the quark-antiquark pair emanating from 
the weak decay vertex to form the other light meson, this gives rise 
to the third contribution to the factorization formula (\ref{fff}). 
This further contribution, called 
``hard-spectator interaction'', involves three LCDAs, including 
the one of the $B$ meson and resembles the expressions that 
appear in the theory of exclusive form factors at large 
momentum transfer~\cite{Lepage:1980fj,Efremov:1979qk}. 

The above discussion reflects the theoretical understanding of 
QCD factorization in exclusive $B$ decays as of around 2000. The 
present article summarizes results from a long-term project 
on soft-collinear factorization in $B$-meson decays started 
in 2003. The three following sections are devoted to three different  
strains of research. The first covers the further conceptual 
development of QCD factorization which centres around (a) factorization 
of the $B\to \pi$ form factor and hard-spectator scattering, 
(b) a calculable example of factorization of endpoint divergences, 
and (c) factorization at sub-leading power in $\Lambda/m_b$ for 
the semi-inclusive semi-leptonic decays $B \to X_u\ell\nu$.
The hard-scattering kernels $T_{ij}^{\rm I}(u)$ and $T_i^{\rm II}(\xi,u,v)$
were computed in~\cite{Beneke:1999br,Beneke:2000ry} at $\mathcal{O}
(\alpha_s)$, which corresponds to the one-loop and tree approximation, 
respectively. Since then most of the required calculations 
for the next order have been done and will be discussed in the 
second part. One of the most remarkable implications of (\ref{fff}), 
(\ref{fff2}) is that the strong phases, which must be present to make direct 
CP violation observable, are either $\mathcal{O}(\alpha_s)$ from 
loops at leading power, or $\mathcal{O}(\Lambda/m_b)$ power-suppressed. 
The next order therefore also yields the first QCD correction to the 
direct CP asymmetries, which is usually required to obtain a reliable 
result. Finally, some aspects of the phenomenology of the factorization 
approach for charmless and other exclusive decays will be very briefly 
presented in the third section. 

I conclude this introduction with an apology. The circumstances that 
require the writing of this article imply that it focuses on my own work. 
It does not do justice to that of many others and does not substitute 
a review of the subject that remains yet to be written.

\section{Conceptual development of QCD factorization}

The further conceptual development of QCD factorization in $B$ decays 
has greatly benefited from the development of soft-collinear 
effective theory (SCET)
\cite{Bauer:2000ew,Bauer:2000yr,Bauer:2001yt,Beneke:2002ph,Beneke:2002ni}, 
the effective Lagrangian formulation of diagrammatic factorization. 
In turn, the desire to better understand the factorization of 
exclusive heavy-quark decays has been instrumental in the early 
development of SCET. For example, the essence of the diagrammatic two-loop 
analysis \cite{Beneke:2000ry} that demonstrated that the structure of 
infrared singularities in the quark representation of the $B\to D\pi$ 
amplitude is consistent with (\ref{fff2}), is elegantly reproduced 
by the decoupling of soft gluons from the leading-power collinear SCET 
Lagrangian through a field redefinition by a soft Wilson line, and the 
subsequent cancellation of the Wilson lines from the colour-singlet current 
that overlaps with the pion state~\cite{Bauer:2001cu}. This proves 
factorization to all orders, provided that SCET reproduces all the 
relevant infrared degrees of freedom, which is the standard assumption.
In the following, however, I focus on hard-spectator scattering, 
where not all the factorization properties are evident in (\ref{fff}), 
since two perturbative scales $m_b$ and $\sqrt{m_b\Lambda}$ are 
hidden in the scattering kernel $T_i^{\rm II}(\xi,u,v)$. It turns out that 
the main issues can already be understood from the factorization of 
heavy-to-light form factors, so I turn to these first.

\subsection{Heavy-to-light form factors}

\begin{figure}
\vskip0.2cm
\begin{center}
\includegraphics[width=2cm]{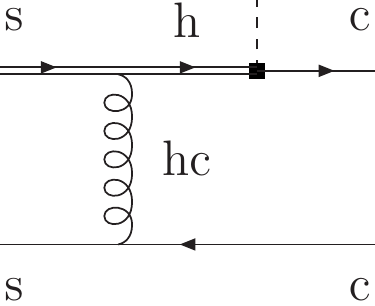}
\hskip0.5cm
\includegraphics[width=2cm]{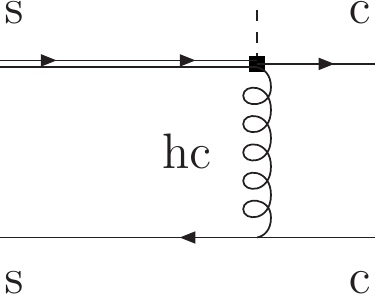}
\hskip0.5cm
\includegraphics[width=2cm]{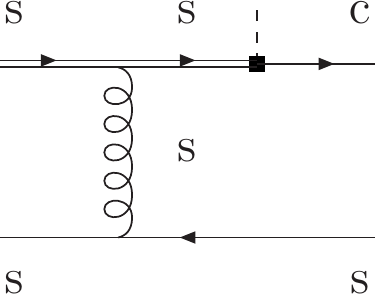}
\end{center}
\vspace*{-0.3cm}
\caption{Tree diagrams for the $B\to \pi$ form factor. Left and middle: 
the pion is produced in a symmetric configuration. In the middle 
diagram hard fluctuations have already been integrated out. Right: when 
the pion is produced in an asymmetric configuration, the diagram 
is purely soft.}
\label{fig:fftree}
\end{figure}

The heavy-to-light form factors parametrize matrix elements of the 
form $\langle M(p^\prime)|\bar q \Gamma b|\bar B(p)\rangle$, where 
$\Gamma$ denotes a Dirac matrix, and we are interested here in the 
region of large momentum transfer of $\mathcal{O}(m_b)$ to the light 
meson $M$. The physics contained in this matrix element is surprisingly 
rich and complex. First attempts compute the large-recoil 
form factor in terms of 
LCDAs date back to \cite{Szczepaniak:1990dt} and are based on the 
assumption that the light meson is produced in a configuration where 
the quark and antiquark constituents both carry large momentum, 
which requires the exchange of an energetic gluon with large virtuality 
$m_b \Lambda$ (left diagram in Figure~\ref{fig:fftree}). However, the 
calculation exhibits an endpoint divergence, which invalidates this 
assumption. The opposite starting point was adopted 
in \cite{Charles:1998dr}, where it was shown that the three (seven) 
independent scalar form factors that parametrize the transition of 
a pseudo-scalar $B$ meson to a pseudo-scalar (vector) light meson 
can be expressed in terms of only a single (of only two) function(s), 
if the transition is dominated by soft interactions. This corresponds 
to the right diagram in Figure~\ref{fig:fftree} and implies that the 
light meson is always produced in an asymmetric configuration, which 
though improbable is favoured by the presence of the soft remnant of 
the $B$-meson after the heavy-quark decay. Ref.~\cite{Beneke:2000wa} 
showed that the symmetry relations that leads to the reduction of 
the number of independent form factors are broken by perturbatively 
calculable corrections, and that the endpoint divergences that appeared 
in the hard-scattering approach could be absorbed into the 
soft form factors. This led to the conjecture that at leading power 
in the $\Lambda/m_b$ expansion the scalar form factors factorize 
as 
\begin{eqnarray}
F_i^{BM}(q^2) = C_i(q^2) \, \xi_{BM}(q^2) + \phi_B \otimes T_i(q^2) \otimes
\phi_M
\label{BMtheorem}
\end{eqnarray}
at large recoil~\cite{Beneke:2000wa}. Here $\otimes$ denotes a convolution 
in momentum fraction, and $\xi_{BM}(q^2)$ refers to the soft form factor 
that satisfies the symmetry relations.

\begin{table}[t]
\caption{Scaling of momentum modes relevant to $B \to M$ form factors. }
\label{table1}     
\begin{center}
\begin{tabular}{llc}
\hline\noalign{\smallskip}
$(n_+p,p_\perp,n_-p)$ & terminology & virtuality  \\
\noalign{\smallskip}\hline\noalign{\smallskip}
$(1,1,1)$ & hard & 1 \\
$(1,\lambda,\lambda^2)$ & hard-collinear & $\lambda^2$ \\
\noalign{\smallskip}\hline
&&\\[-0.3cm]
$(\lambda^2,\lambda^2,\lambda^2)$ & soft & $\lambda^4$ \\
$(1,\lambda^2,\lambda^4)$ & collinear & $\lambda^4$ \\
\noalign{\smallskip}\hline
\end{tabular}
\end{center}
\end{table}

With the advent of SCET the heavy-to-light form factors were extensively 
studied \cite{Bauer:2000yr,Beneke:2002ph,Chay:2002vy,Bauer:2002aj,Beneke:2003pa,Lange:2003pk} resulting in a ``proof'' of (\ref{BMtheorem}) in 
\cite{Beneke:2003pa}.  We first note that there are two large scales in 
the problem. Their existence can already be seen in the left tree diagram in 
Figure~\ref{fig:fftree}. In order to convert the soft spectator quark 
with momentum $\Lambda$ into a energetic light quark, the exchanged 
gluon must have virtuality $m_b\Lambda$. The heavy-quark line that 
connects the gluon vertex to the external weak vertex, however, is 
off-shell by the larger amount $m_b^2$. 
Factorization should therefore proceed in 
two steps through the construction of two effective theories, above 
and below the scale $\sqrt{m_b \Lambda}$~\cite{Bauer:2002aj}. When the 
off-shell heavy quark in the left diagram is integrated out, the effective 
interaction looks as in the middle diagram of Figure~\ref{fig:fftree}, 
which suggests that higher-dimensional operators with additional gluon 
fields must be relevant~\cite{Bauer:2002aj}.

Let the four-momentum of the light meson be $p^\prime = E n_-$, where 
$n_\pm$ define two light-like momenta with $n_+ \cdot n_- =2$, and 
$\lambda = \sqrt{\Lambda/m_b}$ defines the small parameter that organizes the 
power-counting in SCET. The relevant momentum modes are summarized 
in Table~\ref{table1}. In the $B$-meson rest frame, the $B$ meson 
consists of a heavy quark with soft residual momentum and soft 
constituents, while the light meson is made of collinear partons. 
The virtuality of soft and collinear modes is $\Lambda^2$ and 
non-perturbative. In contrast, the hard and hard-collinear modes 
can be integrated out perturbatively one after the other. Each step 
defines a new effective theory, called SCET$_{\rm I}$ and SCET$_{\rm II}$ 
in~\cite{Bauer:2002aj}, and SCET(hc,c,s) and SCET(c,s) 
in \cite{Beneke:2003pa}. In the following I summarize the key results 
in each step, referring to the original papers for many important details. 
To be definite, $M$ will be assumed to be a pion.

In the first step, the hard modes are integrated out. The SCET 
Lagrangian is not renormalized to any order 
\cite{Beneke:2002ph}, but the QCD currents 
$\psi \Gamma Q$   must be matched to SCET$_{\rm I}$. The relevant 
SCET$_{\rm I}$ current operators have field 
content~\cite{Bauer:2002aj}
\begin{equation}  
\label{3ops} 
\bar{\xi}_{c} h_v,\quad  
\bar{\xi}_{c} A_{\perp{ hc}}^{\mu} h_v,
\end{equation} 
where here $c$ can be $c$ or $hc$. The two operators have different $\lambda$ 
scaling in SCET$_{\rm I}$, but nevertheless both contribute to the 
leading power in $\Lambda/m_b$ to the form factor. The reason for this is 
that SCET$_{\rm I}$ contains modes of different virtualities, so that 
the scaling of the matrix elements of (\ref{3ops}) is determined only after 
an analysis of the time-ordered products of the currents with interactions 
from the SCET$_{\rm I}$ Lagrangian. Power-suppressed interactions are 
needed to obtain a term with non-vanishing overlap with the 
pion and $B$-meson state, and the analysis shows that both 
types of operators contribute at order 
$(\Lambda/m_b)^{3/2}$~\cite{Beneke:2003pa}.

Since the (hard-) collinear modes have $n_+\cdot p\sim \mathcal{O}(m_b)$, 
SCET operators are generally non-local and have to be dressed with 
collinear Wilson lines $W_c$ to make them gauge-invariant. The precise 
matching equation is 
\begin{widetext}
\begin{equation}
\label{match1}
    (\bar{\psi}\hspace*{0.03cm}\Gamma_i \hspace*{0.03cm}Q)(0)=
     \int d\hat{s}\,\widetilde{C}^{(A0)}_{i}(\hat{s})\,
    O^{(A0)}(s;0) +\int d\hat{s}_1 d\hat{s}_2\,
    \widetilde{C}^{(B1)}_{i\mu}(\hat{s}_1,\hat{s}_2)\,
    O^{(B1)\mu}(s_1,s_2;0)+\cdots,
\end{equation}
where 
\begin{equation}
   O^{(A0)}(s;x) \equiv (\bar{\xi} W_c)(x+sn_+)
    h_v(x_-) \equiv (\bar{\xi} W_c)_s h_v,\qquad 
   O_{\mu}^{(B1)}(s_1,s_2;x) \equiv \frac{1}{m_b} \,
    (\bar{\xi}W_c)_{s_1} (W_c^\dag i
    D_{\perp c\mu}W_c)_{s_2} h_v
\label{opbasis}
\end{equation}
correspond to the two operators in (\ref{3ops}), 
and $x_-^\mu = (n_+\cdot x) n_-^\mu/2$, 
$\hat{s}_i\equiv s_i m_b/n_- v$. One of the $s$-integrations 
can be removed upon taking the matrix elements by using translation 
invariance. We then define two leading-power 
SCET$_{\rm I}$ pion form factors through   
\begin{equation}
\hspace*{2cm} \langle \pi(p^\prime)| (\bar\xi W_c)h_v|\bar B_v\rangle
 = 2 E\,\xi_{B\pi}(E), \qquad
\end{equation}
\begin{equation}
\hspace*{2cm} 
\langle \pi(p^\prime)| \frac{1}{m_b} (\bar\xi W_c)(W_c^\dagger \,i\!
\not\!\!{D}_{c\perp} W_c)(rn_+) h_v|\bar B_v\rangle
 = 2 E \int d\tau \,e^{i\,2E\tau r}\,\Xi_{B\pi}(\tau,E).
\label{XiP}
\end{equation}
Here $|\bar B_v\rangle$ denotes the $\bar B$-meson state in the 
static limit normalized to $2 m_B$. At this step, the three 
pion form factors 
can be represented as
\begin{eqnarray}
\label{ff2} 
\hspace*{2cm} F_i^{B\pi} (E)\,=\,
C_i(E)\,\xi_{B\pi}(E)\,+\int d\tau\,C^{(B1)}_i(E,\tau)\,
\Xi_{B\pi}(\tau,E), \qquad i=+,0,T.
\end{eqnarray}
\end{widetext}
\noindent The point to note here is that the three
form factors of the $B\to \pi$ transition can be expressed 
in terms of a single form factor $\xi_{B\pi}(E)$ and another 
non-local form factor $\Xi_{B\pi}(\tau,E)$. A 
number of ``symmetry'' relations between form factors emerges 
already at this stage.

The first term of (\ref{ff2}) is already in the form of (\ref{BMtheorem}) 
and is often associated with the ``soft form factor''. While it certainly 
includes the soft overlap contribution, i.e. the contribution where the 
pion is formed in the asymmetric configuration (see the right diagram 
of Figure~\ref{fig:fftree}), it is defined as a SCET$_{\rm I}$ matrix 
elements still containing hard-collinear short-distance effects. The matching 
of the operator $(\bar{\xi} W_c)_s h_v$ to SCET$_{\rm II}$ by integrating 
out the hard-collinear effects has been studied in 
\cite{Beneke:2003pa,Lange:2003pk}, but has so far been unsuccessful. The 
problem arises from the absence of a consistent interpretation of 
endpoint divergences that appear in the convolution integrals of the 
matching coefficients, which point to missing degrees of freedom. The 
analysis of \cite{Beneke:2003pa} shows that the structure must be 
complicated, since three-particle LCDAs of the $B$-meson and pion appear 
at leading power. $\xi_{B\pi}(E)$ is therefore treated as an unknown, 
non-perturbative function. Fortunately, there is only one instead of the 
original three (two instead of seven for vector mesons).

Yet (\ref{ff2}) would be of little use, if the second term containing 
an unknown form factor $\Xi_{B\pi}(\tau,E)$ of {\em two} variables could 
not be simplified. In the second matching step, one therefore integrates 
out the hard-collinear scale $\sqrt{m_b \Lambda}$. Note that this step 
is only applied to the second term and not the first for the reasons 
mentioned above. A power-counting analysis \cite{Beneke:2003pa} 
shows that the operator 
$O_\mu^{(B1)} \sim \bar\xi_cA^\mu_{\perp  hc} h_v$ matches only on 
four-quark operators of 
the form $(\bar{q}_s h_v) (\bar\xi_c\xi_c)$ with no additional 
fields or derivatives at leading power in the $\lambda$ expansion. 
The explicit operator matching relation reads 
\begin{widetext}
\begin{eqnarray}
&&  
2 E \int \frac{d r}{2\pi}\,e^{-i\,2 E \tau r}\, (\bar\xi
W_c)(0) (W_c^\dagger \,i\!
\not\!\!{D}_{c\perp} W_c)(r n_+)h_v(0)
\nonumber  \\
&& = \, \int d\omega dv \,J(\tau;v,
\ln (E\omega/\mu^2)) \,\Big[(\bar\xi W_c)(s n_+)\frac{\not\!\!n_+}{2}\gamma_5
(W_c^\dagger\xi)(0)\Big]_{\rm FT} \Big[(\bar q_s Y_s)(t
n_-)\frac{\not\!\!n_-}{2}\gamma_5 (Y_s^\dagger h_v)(0)\Big]_{\rm
FT}\,+\,\ldots
\label{matchrel}
\end{eqnarray}
with
\begin{eqnarray}
Q(v) &\equiv& \Big[(\bar\xi W_c)(s n_+)\frac{\not\!\!n_+}{2}\gamma_5
(W_c^\dagger\xi)(0)\Big]_{\rm FT} 
=  \frac{n_+
p^\prime}{2\pi}\int ds\,e^{-is v n_+ p^\prime}\, (\bar\xi W_c)(s
n_+)\frac{\not\!\!n_+}{2}\gamma_5(W_c^\dagger\xi)(0),
\\[0.2cm]
P(\omega) &\equiv & 
\Big[(\bar q_s Y_s)(t
n_-)\frac{\not\!\!n_-}{2}\gamma_5 (Y_s^\dagger h_v)(0)\Big]_{\rm FT}
= \frac{1}{2\pi}\int dt\,e^{it \omega}\,
(\bar q_s Y_s)(t n_-)\frac{\not\!\!n_-}{2}\gamma_5(Y_s^\dagger h_v)(0),
\label{Pdef}
\end{eqnarray} 
\end{widetext}
\noindent and $J(\tau;v,\ln (E\omega/\mu^2))$ the hard-collinear matching 
coefficient, which can be computed perturbatively. (The middle 
diagram in Figure~\ref{fig:fftree} is one of the two tree diagrams 
that contribute to $J$.) $Y_s$ denotes a soft Wilson line. The Wilson 
lines $W_c$, $Y_s$ ensure the gauge-invariance of the light-cone 
operators.

In the SCET$_{\rm II}$ Lagrangian there are no interactions between 
collinear and soft fields, since the sum of a collinear and soft momentum 
has hard-collinear virtuality, which is already integrated out. The 
SCET$_{\rm II}$ matrix element of a product of collinear and soft 
fields as it appears on the right-hand side of (\ref{matchrel}) always 
falls apart into separate collinear and soft matrix elements. In the 
present case, these define the LCDA
\begin{equation}
\langle \pi(p^\prime)|Q(v)|0\rangle =
-if_\pi E\,\phi_\pi(v)
\end{equation}
for the pion and 
\begin{equation} 
\langle 0|P(\omega)|\bar B_v\rangle = 
 \frac{i \hat{f}_B m_B}{2}\,\phi_{B+}(\omega)
\label{blcda}
\end{equation}
for the $B$-meson \cite{Beneke:2000wa,Grozin:1996pq}. Thus the 
four-fermion operator factorizes into simpler matrix elements upon 
integrating out the hard-collinear scale and the non-local form factor 
can be written as \cite{Beneke:2003pa}
\begin{widetext}
\begin{equation}
\hskip2.5cm 
\Xi_{B\pi}(\tau,E) = \frac{m_B}{4m_b}\,\int_0^\infty d\omega\,\int_0^1 dv\,
J(\tau;v,\ln (E\omega/\mu^2))\,\hat{f}_B\phi_{B+}(\omega)f_\pi\phi_\pi(v).
\label{b1match}
\end{equation}
\end{widetext}
\noindent
An important point needs to be made here. Due to the absence of 
soft-collinear interactions in the SCET$_{\rm II}$ Lagrangian {\em any} 
operator at any order in the $\lambda$ expansion 
formally factorizes into a convolution of a soft and a collinear 
function, but the result is not always correct, since the convolution 
integrals often diverge at the endpoints --- 
see the discussion of $\xi_{B\pi}(E)$ 
above. The final part of the factorization ``proof'' therefore consists 
of showing that the convolution integrals in (\ref{b1match}) converge. 
The argument given in \cite{Beneke:2003pa} shows that $J$ can depend 
on $\omega$ only through terms of the form $1/\omega\times 
\ln^n(E\omega/\mu^2)$, which together with the endpoint behaviour of the 
$B$-meson LCDA guarantees the convergence of the $\omega$-integral. 
From this the convergence of the $v$-integral is inferred, since an 
endpoint divergence in a collinear integral must always have a correspondence 
in a soft convolution.

The subtleties of soft-collinear factorization  are 
related to the factorization of modes with equal rather than hierarchical 
virtuality (\cite{Beneke:2003pa}, Figure~2), which are distinguished by 
the scaling of a longitudinal light-cone momentum  component 
rather than transverse momentum. The factorization of such 
momentum regions usually leads to divergences that are not regulated 
in dimensional regularization. Other schemes that work such as analytic 
regularization whereby one raises certain propagators to non-integer 
powers $1/[-k^2]^{1+a}$ violate some of the longitudinal 
boost symmetries of the classical 
Lagrangian, which constrain the dependence of coefficient functions and 
operators on light-cone momentum components in the naive factorization 
formula. The absence of a regulator that preserves all the properties of 
the classical theory at the quantum level implies that SCET$_{\rm II}$ 
factorization is generally ``anomalous'' \cite{Beneke:2005dubna} 
and valid only in special 
cases, such as the form factor $\Xi_{B\pi}(\tau,E)$ discussed above. 
To my knowledge, a general argument when SCET$_{\rm II}$ factorization 
is {\em not} anomalous is not available. These difficulties do not affect 
the first matching step where hard modes are integrated out. At least, 
I do not know of an example for which the convolution in the 
hard-collinear momentum fraction ($\tau$ above) is divergent or 
dimensional regularization fails in matching the full theory to 
SCET$_{\rm I}$.

Returning to the heavy-to-light form factor we plug (\ref{b1match}) 
into (\ref{ff2}) and obtain the conjectured factorization formula 
(\ref{BMtheorem}), provided we identify the kernel $T_i(q^2)$ of the 
spectator-scattering term with 
\begin{widetext}
\begin{equation}
\hskip2.5cm T_i(E;v,\ln \omega) = \frac{m_B}{4m_b}
\int d\tau\,C^{(B1)}_i(E,\tau)\,\hat{f}_B f_\pi 
J(\tau;v,\ln (E\omega/\mu^2))\,,
\label{eq:Tform}
\end{equation}
\end{widetext}
\noindent
or $T_i = C^{(B1)}_i \otimes J$ in short. The hard-scattering kernels of 
spectator scattering are themselves a convolution of a hard and a 
hard-collinear function~\cite{Bauer:2002aj}. The one-loop corrections 
to $C^{(B1)}_i$ \cite{Beneke:2004rc,Becher:2004kk} and $J$ \cite{Hill:2004if,Beneke:2005gs} have both been calculated, and the explicit 
results confirm the general factorization formula, including the 
arguments supporting the convergence of the convolution integrals. 
We therefore conclude that the three independent $B\to\pi$ form factors 
can be expressed in terms of a single form factor $\xi_{B\pi}$ 
plus corrections 
that begin at $\mathcal{O}(\alpha_s(\sqrt{m_b\Lambda}))$ and can be 
expressed in terms of the universal, process-independent LCDAs of the mesons.

\subsection{Spectator scattering in hadronic decays}

I have discussed the heavy-to-light form factors at length, since after 
these preparations the extension to charmless $B$ decays is relatively 
simple. Charmless decays were first considered in SCET 
in \cite{Chay:2003zp,Chay:2003ju}. After the development of the theory 
of spectator scattering for the form-factor, 
the derivation of the QCD factorization formula within the SCET framework 
was completed in~\cite{Bauer:2004tj}. 

Instead of current operators one needs to match the operators present 
in the weak effective Hamiltonian. As an example, consider the 
contribution of $Q = [\bar u_b \gamma^\mu (1 -\gamma_5) b_a]
[\bar d_a \gamma_\mu (1 - \gamma_5) u_b]$ to the decay $\bar B_d 
\to \pi^+\pi^-$ ($a,b$ are colour indices). 
The extra $\pi^-$ relative to the $B\to \pi$ transition 
has large momentum $q$ in the direction opposite to 
$\pi^+$. The SCET Lagrangian is therefore extended by a second collinear 
sector with fields $\chi$, $A_{c2}$ in addition to 
$\xi$, $A_{c1}$, satisfying $\not\!\!n_-\xi=0$ and $\not \!\!n_+\chi=0$.
Analogous to (\ref{match1}) the operator $Q$ matches to the 
expression
\begin{equation}
Q = \!\int \!d\hat{t}\,\tilde{T}^{\rm I}(\hat t) O^{\rm I}(t) + \!
\int \!d\hat{t}d\hat{s}\,\tilde{H}^{\rm II}(\hat t,\hat s)\OII(t,s)
\label{match1nonlep}
\end{equation}
at leading power after integrating out the hard modes, where the 
two operator structures are now given by
\begin{widetext}
\begin{eqnarray}
O^{\rm I}(t) &\!=\!& 
  (\bar \chi\Wcp) (t n_-) \nmh (1 - \gamma_5) (\Wcp^\dag \chi) \;
        \Big[ \tilde C^{(A0)}_{f_+} \,(\xib W_{c1}) 
        \npslash (1 - \gamma_5) h_v 
\nonumber \\*
&&\hspace*{1cm} 
  - \frac{1}{m_b} \int d\hat{s}\, \tilde C^{(B1)}_{f_+}(\hat{s})\,
    (\bar \xi W_{c1}) \npslash [W^\dagger_{c1}i 
        \Slash{D}_{\perp c1} W_{c1}](s n_+)
                (1 + \gamma_5) h_v \Big] ,
\label{opsninlep1}\\[0.2cm]
\OII(t,s) &\!=\!& \frac{1}{m_b} \Big[
   (\bar \chi\Wcp)(t n_-) \nmh (1 - \gamma_5) (\Wcp^\dag\chi)\Big] 
\;\Big[(\bar \xi W_{c1}) \nph [W^\dagger_{c1}i \Slash{D}_{\perp c1}
                 W_{c1}](s n_+) (1 + \gamma_5) h_v\Big].
\label{opsnonlep2}
\end{eqnarray}
\end{widetext}
\noindent
It might appear that a final state with two pions should create a major 
problem, since the physics of charmless decays is obviously more complicated 
than that of the form factor. However, the two collinear sectors are 
already decoupled after the first matching step to SCET$_{\rm I}$, since 
the sum of a c1 and c2 momentum necessarily has hard virtuality  
$\mathcal{O}(m_b^2)$. Thus there can be no coupling of the two modes 
in the SCET$_{\rm I}$ Lagrangian, and hence the $\langle \pi^-\pi^+|\ldots
|\bar B\rangle$ matrix element of the operators (\ref{opsninlep1}), 
(\ref{opsnonlep2}) factorizes already at the hard-collinear 
scale into a $\langle \pi^-|\ldots
|0\rangle$ matrix element of the c2 fields and the 
$\langle\pi^+|\ldots |\bar B\rangle$ matrix element of the remainder 
after decoupling the soft gluons from the c2 fields by the soft Wilson-line 
field redefinition in the same way as for $\bar B_d\to D^+\pi^-$. 
The former simply gives the pion LCDA $\phi_{M_2}(u)$ in (\ref{fff}).
The latter matrix elements are the same as those that appear in the 
factorization of the $B\to \pi$ form factors. Following \cite{Beneke:2005vv}, 
the square bracket in (\ref{opsninlep1}) has been defined such that its 
matrix element coincides with the full QCD $B\to\pi$ form factor. 
This yields the first term in (\ref{fff}).
The matrix element of the second square bracket in (\ref{opsnonlep2}) 
can be related to $\Xi_{B\pi}$ defined in (\ref{XiP}), so (\ref{b1match}) 
can also be used here. Introducing the Fourier transform
\begin{equation}
H^{\rm II}(u,v) = \int d\hat{t}d\hat{s}\,
e^{i (u\hat{t}+(1-v)\hat {s})}\,
\tilde{H}^{\rm II}(\hat t,\hat s)
\label{eq:Huv}
\end{equation}
of the hard-matching coefficient of the operator $\OII(t,s)$ associated 
with the spectator scattering, where $u$, $v$ correspond to collinear 
momentum fractions,  
one finds that the kernel 
$T^{\rm II}(\omega,u,v)$ in (\ref{fff}) is given by the convolution 
\begin{widetext}
\begin{equation}
\hspace*{2.5cm}
T^{\rm II}(\omega,u,v) = -\frac{m_B}{8 m_b}
\int_0^1 \!dz \,H^{\rm II}(u,z)\,J(1-z;v,\omega).
\label{t2}
\end{equation}
Here $J$ is the same hard-collinear matching coefficient that 
appeared in the factorization of the form factor.
This reproduces the QCD factorization formula (\ref{fff}) for 
charmless decays in the SCET framework. 
\end{widetext}
\noindent 

The SCET derivation yields several
new insights. First, the spectator-scattering kernel is a convolution 
$T^{\rm II} = H^{\rm II} \otimes J$ of a hard and hard-collinear 
function, and precise operator definitions can be given, which 
are important for consistent higher-order calculations 
(see following section). 
Second, the decoupling of the second light meson at the hard-collinear 
scale implies that the perturbative strong interaction phases 
originate from the imaginary part of hard loops, and never from 
hard-collinear loops. Third, the upper limit of the $\xi$-integral 
in (\ref{fff}) should be replaced by an integral over $\omega$ with 
upper limit $\infty$ instead of 1. The variable 
$\xi = \omega/m_B$ corresponds to a light-cone momentum fraction of the 
spectator quark in the $B$ meson. However, the LCDA $\phi_B(\omega)$ 
is defined in SCET$_{\rm II}$, which contains a static quark field 
$h_v$ of formally infinite mass. Relative to this the variable 
$\omega$ can take any value. The one-loop renormalization of 
$\phi_B(\omega)$ \cite{Lange:2003ff} shows that a perturbative tail 
is generated that ranges to $\omega = \infty$, unless the LCDA 
is defined in a cut-off scheme. Since the characteristic scale 
is $\omega \sim \Lambda$, the $B$-meson 
LDCA can appear in leading-power factorization formulae only in the 
form of the inverse moments
\begin{equation} 
\sigma_n(\mu) \equiv \lambda_B(\mu) \int_0^\infty \frac{d\omega}{\omega} 
\,\phi_{B+}(\omega,\mu)\,\ln^n \frac{\mu_0}{\omega},
\label{BLCDAmom}
\end{equation}
where the most important moment $\lambda_B(\mu)$ is defined by 
$\sigma_0(\mu)\equiv 1$. At order $\mathcal{O}(\alpha_s^k)$, up to 
$n = 2 k-2$ logarithms can appear, so the number of parameters related to 
the $B$-meson LCDA is limited to a few in practical calculations. 

\subsection{Variants of factorization for hadronic decays}

Several variations of factorization as described above have been 
advocated. The PQCD framework \cite{Keum:2000wi,Lu:2000em} assumes 
that the $B$-meson transition 
form factors $F_i^{B M_1}(0)$ 
are also dominated by short-distance physics and factorize into 
light-cone distribution amplitudes. Both terms in (\ref{fff}) 
can then be combined to 
\begin{equation}
\langle M_1 M_2|Q_i|\bar B\rangle =
T^{\rm PQCD}\star \phi_B \star \phi_{M_1}\star 
\phi_{M_2}.
\label{ffPQCD}
\end{equation}
The implicit assumption that the heavy-to-light form factors do not 
receive any soft long-distance contribution (due to Sudakov suppression) 
seems difficult to justify to me, and 
certainly contradicts the picture developed above. At the technical level, 
the endpoint divergences are regularized by keeping the intrinsic 
transverse momentum of the meson constituents. There is a larger 
sensitivity to perturbative corrections at low scales, where the strong 
coupling is large and perturbation theory potentially unreliable. 
From a phenomenological perspective, PQCD needs fewer non-perturbative 
input parameters, but there is a larger dependence on unknown 
light-cone distribution amplitudes. A principal difference between 
the PQCD and the QCD factorization approach described above is the relative 
importance of the weak-annihilation mechanism and the generation of 
strong rescattering phases. In QCD factorization the phases arise at the 
scale $m_b$ from loop diagrams that 
have yet to be included in the PQCD approach, and from a model 
for power-suppressed weak annihilation, whereas in the most widely used 
implementation of PQCD the strong phases originate only from a 
weak-annihilation tree diagram. 
A recent attempt to compute radiative corrections in the PQCD 
approach resulted in 
uncancelled infrared divergences \cite{Li:2009wba}, which led to 
the introduction of final-state specific non-perturbative parameters, 
in violation of factorization. Since a clear power-counting scheme 
in $\Lambda/m_b$ has never been established for the PQCD formula 
(\ref{ffPQCD}), this is not necessarily in contradiction with 
factorization at leading power in $\Lambda/m_b$. However, it 
prevents a systematic improvement of PQCD 
calculations beyond tree-level.

Ref.~\cite{Bauer:2004tj} proposed a phenomenological implementation of 
factorization\footnote{Sometimes called the ``SCET approach'', which 
I find misleading, since the theoretical basis of diagrammatic 
QCD factorization and SCET factorization is exactly the same. See the 
derivation of (\ref{fff}) in the previous subsection.} 
that differs in two important respects from the BBNS 
approach~\cite{Beneke:1999br,Beneke:2000ry,Beneke:2001ev}. 
First, perturbation theory at the intermediate scale $\sqrt{m_b\Lambda}$ is 
avoided by not factorizing the spectator-scattering term into a hard and 
jet function. Eq.~(\ref{fff}) then takes the form 
\begin{eqnarray}
\langle M_1 M_2|Q_i|\bar B\rangle &\!=\!& F^{BM_1}\, 
{T_i^{\rm I}} \star \phi_{M_2} 
\nonumber\\ 
&&+\,  
\Xi^{BM_1}\star {H_i^{\rm II}}\star \phi_{M_2}\,,
\label{ff1SCET}
\end{eqnarray}
which is simply the matrix element of the SCET$_{\rm I}$ representation 
(\ref{match1nonlep}) of the four-quark operator of the weak effective 
Hamiltonian. Second, penguin diagrams with charm loops, the so-called 
``charming penguins'' \cite{Ciuchini:2001gv}, are claimed to be 
non-factorizable, hence non-perturbative at leading order in the 
$\Lambda/m_b$ expansion. From the phenomenological perspective the principal 
difference to the BBNS approach concerns again the generation of 
strong interaction phases. Since the non-local form factor is 
unknown, Eq.~(\ref{ff1SCET}) can be used in practice only at tree level, 
hence the matrix elements have no imaginary parts. The only exception is the 
matrix element corresponding to the charm penguin amplitude, which is 
considered as an unknown complex number (one for each final state, 
to be fitted to data), and 
therefore represents the only source of direct CP violation. 
Weak-annihilation effects are neglected, since they are power-suppressed. 
A critique of this approach is given in \cite{Beneke:2004bn}. From the 
theoretical perspective of factorization, the most important issue 
raised is the question whether the penguin loops with  
charm factorize or not. The concern is that for massive quark loops 
the non-relativistic energy scale $m_c v^2$ near threshold is of order 
$\Lambda$, and hence there appears to be a sensitivity to non-perturbative 
scales not present for massless quark loops. The suppression of 
sensitivity to this region and the associated issue of quark-hadron 
duality for charm-quark loops has been discussed in \cite{Beneke:2009az}, 
which resolves the issue in favour of factorization.

\subsection{Endpoint divergences in quarkonium final states}

Two-body decays to a charmonium ($M_2$) and a light meson ($M_1$) were 
argued~\cite{Beneke:2000ry} to also factorize according to (\ref{fff}) 
in the non-relativistic (Coulombic) 
limit $m_c v^2 \gg \Lambda$, since the size 
of the charmonium $1/(m_c v)$ is small compared to the wave-length 
of soft gluons, which therefore decouple. However, when the calculation of 
the hard-scattering functions was done for P-wave charmonia 
\cite{Song:2002mh}, uncancelled 
divergences appeared, including an endpoint divergence in spectator 
scattering, which contradicted the expectation of factorization.
In this section I briefly review the result of \cite{Beneke:2008pi}, 
which explains the origin of the divergences and resurrects factorization 
at leading order in $\Lambda/m_b$. The main conceptual interest in this 
result derives from the fact that (to my knowledge) it constitutes the 
only example of a consistent factorization of an endpoint divergence 
in terms of precisely defined operator matrix elements.

An important property of final states with charmonium when the latter is 
assumed to be a non-relativistic bound state is that there exist two sets 
of low-energy scales, $m_c v, m_c v^2$ related to the non-relativistic 
expansion, and
$\sqrt{m_b\Lambda}$, $\Lambda$, related to the collinear expansion
and the strong-interaction scale. In the following we assume that 
$m_c \gg \sqrt{m_b\Lambda} \gg m_c v, m_c v^2\gg \Lambda$. 

Integrating out the scales $m_b$, $m_c$ leads from QCD to an 
effective theory that combines SCET$_{\rm I}$ 
with the non-relativistic effective 
theory of QCD for the charm quarks. A QCD operator such as 
$Q = [\bar c_b \gamma^\mu (1 -\gamma_5) b_a]
[\bar s_a \gamma_\mu (1 - \gamma_5) c_b]$, which contributes to the 
decay $\bar B_d \to \chi_{cJ} \bar{K}^0$, matches to 
SCET$_{\rm I}\otimes\,$NRQCD operators which have exactly the same 
form as (\ref{opsninlep1}), (\ref{opsnonlep2}) except that the part 
referring to the second light meson has to be replaced by a non-relativistic 
P-wave colour-singlet rather than a light-cone operator:  
\begin{widetext}
\begin{equation}\hspace*{2cm}
 (\bar \chi\Wcp) (t n_-) \nmh (1 - \gamma_5) (\Wcp^\dag \chi)
\quad\to\quad \mathcal{O}(^{2S+1}\!P_J^{(1)}) \equiv 
\psi_v^{\dagger} \Gamma_\mu \left(-\frac{i}{2}\right)
\stackrel{\leftrightarrow}{D}^\mu_{\top}\chi_v
\end{equation}
\end{widetext}
\noindent 
with $D^\mu_\top = D^\mu - (v\cdot D) v^\mu$ and $\Gamma_\mu$ a 
spin matrix. One now finds that the 
tree-level matching coefficient of the operator $\OII$ (\ref{opsnonlep2}) 
contains terms of the form $m_c^2/(m_b^2 (1-y))$ at tree-level, 
which are absent for massless quarks. Since $\bar y\equiv 1-y$ is the fraction 
of hard-collinear momentum carried by the gluon field in (\ref{opsnonlep2}), 
the divergence at $y=1$ occurs in the region when no momentum is 
transferred to the spectator quark, which is the endpoint region. 
Indeed, one then finds that the hard-spectator amplitude 
contains an integral of the form 
\begin{equation}
\mathcal{A}^{\mbox{{\scriptsize{hard spectator}}}}_
{B\rightarrow H(^{2S+1}P_J) K} \supset
\frac{f_K \hat{f}_B m_B}{m_b \lambda_B}
\int_0^1 d y \,\frac{\phi_{\bar K}(y)}{\bar y^2},
\label{hardspecamp1}
\end{equation}
which diverges at $y=1$.

A crucial point is that the presence of the non-relativistic scales 
allows additional operators to contribute at leading order in $\Lambda/m_b$. 
New and central to the present discussion are colour-octet operators 
such as 
\begin{equation}
\mathcal{O}^{\rm I}_{\perp}(^{3}S_{1}^{(8)})=
\bar{\xi} W_c\gamma_{\perp \mu}(1-\gamma_5) T^A h_w
\, \psi_v^{\dagger}\gamma_{\top}^{\mu} T^A \chi_v.
\label{octetop}
\end{equation}
In the case of charmless decays to two light mesons 
the matrix elements of colour-octet operators can be non-zero
only due to power-suppressed soft-gluon interactions, where soft
means momentum of order $\Lambda$, thus they
can be neglected at leading order in the $\Lambda/m_b$ expansion.
For charmonium, however, the decoupling of gluons with small momentum 
holds only when the momentum is much smaller than $m_c v^2$; 
gluons with momentum $m_c v^2$ contribute
to the octet operator matrix elements even at leading order in  $\Lambda/m_b$.
These contributions are sub-leading in $v$,
but so are the $P$-wave operators due to the extra derivative in
$\mathcal{O}(^{2S+1}P_J^{(1)})$,
hence the gluon-exchange contribution to the S-wave
octet operators is relevant at leading order in the velocity expansion 
to P-wave charmonium production.

Spectator scattering thus consists of hard spectator scattering, where 
the gluon exchanged between the $c\bar c$ system and the spectator 
quark in the $B$ meson has energy of order $m_b$, 
and $\bar y\sim \mathcal{O}(1)$; 
and soft spectator scattering with gluon energy $\mathcal{O}(m_c v^2)$, 
in which case $\bar y\sim \mathcal{O}(v^2)\ll 1$ 
is naturally in the endpoint region. 
Soft spectator scattering leaves the charm quarks close to their mass-shell 
and therefore does not reduce to an effective interaction 
of the form (\ref{opsnonlep2}). It is part of the colour-octet matrix 
element to be computed within the effective theory and is sensitive to the 
charmonium bound-state dynamics. The explicit calculation in the 
Coulombic limit $m_c v^2\gg \Lambda$ gives an expression that contains 
\begin{widetext}
\begin{equation}
\hspace*{1cm}\mathcal{A}^{\mbox{{\scriptsize{soft spectator}}}}_
{B\rightarrow H(^{2S+1}P_J) K} \supset
\frac{f_K \hat{f}_B m_B}{m_b \lambda_B}
\int_0^1 d y \,\phi_{\bar K}(y)\,
\left(\sqrt{-\left(\,\bar y+\frac{2\sqrt{z} E_H}{m_b(1-z)}
\right)}+\frac{\gamma_B}{m_b\sqrt{1-z}}\right)^{-4},
\label{softspecamp1}
\end{equation}
\end{widetext}
\noindent 
where $E_H<0$ is the binding energy of the charmonium state, 
$z=4 m_c^2/m_b^2$ and $\gamma_B = (m_c\alpha_s C_F)/2$ is the inverse 
Bohr radius. We now compare the integral over the kaon 
distribution amplitude to (\ref{hardspecamp1}). While the integrand there 
was applicable to $y$ not near 1 and exhibited a logarithmic endpoint
divergence as $y\to 1$, the integrand of (\ref{softspecamp1}) is appropriate
only to $1-y\sim v^2$, i.e. {\em in} the endpoint region. There is no divergence
here as $y\to 1$. However, for $\bar y\gg v^2$ the integrand has the
same logarithmic behaviour $\int dy\,\phi_{\bar K}(y)/\bar y^2$ as 
does the hard-spectator contribution for $\bar y\ll 1$. 
Eq.~(\ref{hardspecamp1}) is based on approximations that are invalid 
when $\bar y$ is small, which causes the endpoint divergence. 
We can regulate the divergent integral by cutting off the
$y$ integral above $1-\mu$. This corresponds to a hard factorization
scale in the energy of the gluon that connects to the spectator
quark. The spectator-scattering contribution to the colour-octet
matrix element originates precisely from the energy region that
is then cut out in (\ref{hardspecamp1}), and is not valid when 
$\bar y\sim \mathcal{O}(1)$, thus the correct interpretation
of the $y$-integral in (\ref{softspecamp1}) is
$\int_0^1 dy \to \int_{1-\mu}^1 dy$. To combine with  (\ref{hardspecamp1})
we must evaluate the regularized version of (\ref{softspecamp1})
up to terms of order $v^2/\mu$. We then find 
\begin{widetext}
\begin{equation}
\hspace*{1cm}\mathcal{A}^{\mbox{{\scriptsize{soft spectator}}}}_
{B\rightarrow H(^{2S+1}P_J) K} \supset
\frac{f_K \hat{f}_B m_B}{m_b \lambda_B}\,\phi_{\bar K}^\prime(1)
\,\big(-\ln\mu -\ln(1-z) +1+i \pi +\Delta F\big),
\label{softspecamp2}
\end{equation}
\end{widetext}
\noindent 
where $\Delta F$ is a real number  given in \cite{Beneke:2008pi} 
depending on $m_b$, $m_c$, $E_H$ and $\gamma_B$.
The endpoint contribution is proportional to
$\phi_{\bar K}^\prime(1)$, the derivative of the kaon LCDA  
at the endpoint, because the endpoint region is of size $v^2$ rather than
$\Lambda/m_b$, hence it is justified to describe the quarks
in the kaon by collinear quark fields.

There are several remarkable features of this result. First, although 
(\ref{softspecamp1}) comes from a tree-diagram contribution 
to the octet matrix 
element, it carries a sizable absorptive part related to rescattering, 
which originates from the on-shell charm propagator in the integral 
over the charmonium bound-state wave function. Second, the factorization 
scale dependence on $\mu$ cancels exactly with a corresponding 
logarithm in the (regulated) hard spectator contribution 
(\ref{hardspecamp1}). The cancellation occurs as usual between an 
infrared (endpoint) divergence from the coefficient function and an 
ultraviolet (large energy) divergence of the octet matrix element. 
It further occurs between objects that have precise operator definitions 
in the relevant effective field theory.

Unfortunately, these results do not readily generalize to final states 
of two light mesons. In the charmonium case above, there are 
two endpoint regions, $\bar y \sim v^2$ and $\bar y \sim \Lambda/m_b$, 
and we showed the cancellation of the endpoint divergence 
in the former, while still working at leading order in $\Lambda/m_b$.
One difficulty with endpoint factorization at $\bar y \sim \Lambda/m_b$ 
is that a light meson in the asymmetric $q\bar q$ configuration 
cannot be made out of generic collinear and soft modes in SCET$_{\rm II}$, 
since the sum of collinear and soft momenta has virtuality $m_b \Lambda$, 
not $\Lambda^2$. It seems that SCET$_{\rm II}$ is lacking some of the  
relevant degrees of freedom to write down operators that can describe 
the endpoint region and overlap with the light-meson wave function. 
The understanding of this issue seems central to me to further theoretical 
progress on exclusive $B$ decays, which has to address the factorization 
properties of power corrections in $\Lambda/m_b$. The present use of 
factorization in exclusive $B$ decays combines a rigorous theory 
at leading power in $\Lambda/m_b$ with a rudimentary parametrization 
of some power corrections, which introduces a considerable but presently 
unavoidable arbitrariness in some quantities, especially direct CP 
asymmetries, where power-suppressed phases compete with loop-induced ones.

\subsection{Power corrections to semi-inclusive $B$ decay}
\label{sec:semipower}

While a rigorous theory of power-suppressed effects is not available 
for exclusive $B$ decays, there is one important class of processes for 
which soft-collinear factorization has been worked out completely  
including the $\mathcal{O}(\Lambda/m_b)$ terms. These are the 
inclusive semi-leptonic decays $\bar{B}\to X_u \ell\bar\nu$, where 
$X_u$ denotes a hadronic final state without charm, and inclusive 
radiative and electroweak penguin decays $\bar{B}\to X_s\gamma$,   
$\bar{B}\to X_s\ell^+\ell^-$ in the kinematic region where the 
hadronic final-state is collimated into a single jet, which 
carries large energy of $\mathcal{O}(m_b)$, but has small (though 
not too small) invariant mass $m_b \Lambda \ll m_b^2$. Leading-power 
factorization into hard, jet and soft (here called shape) functions 
for these processes was discussed in 
\cite{Bauer:2000yr,Korchemsky:1994jb} in the SCET and diagrammatic 
framework.

The reason why a rigorous treatment of power corrections is possible here 
is the absence of collinear physics at the scale $\Lambda$.
The two-step matching at the hard and hard-collinear scale then 
corresponds  to the sequence ${\rm QCD \to SCET_{\rm I} \to HQET}$ of 
effective theories. The jet functions containing the collinear 
physics at the scale $\sqrt{m_b\Lambda}$ can be computed perturbatively 
in the strong coupling, while the non-perturbative soft physics is 
contained in well-defined HQET matrix elements. There is no other mode 
with virtuality $\Lambda^2$ than the soft mode, when the virtuality 
of the inclusive hadronic final state is $\mathcal{O}(m_b\Lambda)$. 
The effective-theory framework provides a transparent book-keeping 
of the relevant interactions at every step in the calculation, 
including power-suppressed effects, and allows us to  
identify all the relevant HQET operators. The factorization of semi-inclusive 
decays at $\mathcal{O}(\Lambda/m_b)$ has been discussed in
\cite{Lee:2004ja,Bosch:2004cb,Beneke:2004in} at various levels of 
detail. The discussion below outlines the main points following the 
presentation in \cite{Beneke:2004in}, which contains further details 
and the explicit tree-level computation.

The strong interaction effects in semi-leptonic $B$ decays are contained
in the hadronic tensor $W^{\mu\nu}$, related by the optical theorem 
to the imaginary part of the forward scattering amplitude, which we define
as 
\begin{equation}
W^{\mu\nu}= \frac{1}{\pi}\,{\rm Im}\,\langle\bar{B}(v)
|T^{\mu\nu}|\bar{B}(v)\rangle.
\label{hadronic_tensor}
\end{equation}
The  correlator $T^{\mu\nu}$ is the time-ordered product 
of two flavour-changing weak currents
\begin{equation} \label{eq:correlator}
T^{\mu\nu}=i\int d^4 x e^{-iq\cdot x}{\rm
  T}\{J^{\dagger\mu}(x),J^{\nu}(0)\},
\end{equation}
where $q$ is the momentum carried by the outgoing lepton pair
in the $\bar B\to X_u \ell\bar \nu$ decay and 
$J^\mu=\bar{q}\gamma^\mu(1-\gamma_5)b$. 

The matching of the current $J_\mu$ from QCD 
to SCET$_{\rm I}$ is the same as was 
discussed for exclusive decays. Due to rotational invariance in the 
transverse plane, a frame can be chosen such that the hard-collinear 
transverse momenta of order $\lambda= \sqrt{\Lambda/m_b}$ appear only in 
the internal integrations over the momenta in the jet, and since 
the integral over an odd number of transverse momenta either vanishes 
or must be proportional to one of the external soft transverse 
momenta of order $\lambda^2$, the resultant expansion is in 
powers of $\lambda^2\sim 1/m_b$ rather than power of $\lambda$. 
We therefore need to extend (\ref{match1}) 
to second order in the SCET expansion parameter $\lambda$. The basis 
of SCET$_{\rm I}$ heavy-light current operators at $\mathcal{O}(\lambda^2)$ 
contains 15 different operator structures, which are listed in 
\cite{Beneke:2004in}. A few representative examples are 
\begin{eqnarray}
J^{(2)}_{3} &=&(\bar\xi  W_c)_s\,i\overleftarrow{\partial}_{\perp}^\mu\Gamma_j 
(x_\perp D_s h_v)
\nonumber\\
J^{(2)}_8 &=&(\bar\xi W_c)_{s_1} \,i\overleftarrow{\partial}_{\perp}^\mu
[W_c^\dag i D_{\perp c}^\nu W_c]_{s_2}
\Gamma_j h_v 
\end{eqnarray}
and 
\begin{widetext}
\begin{eqnarray}
\hspace*{2cm} 
J^{(2)}_{12} 
&=&(\bar\xi W_c)_{s_1} \Big\{[W_c^\dag i D_{\perp c}^\mu W_c]_{s_2},
 [W_c^\dag i D_{\perp c}^\nu W_c]_{s_3}\Big\} \Gamma_j h_v
\nonumber\\
J^{(2)}_{14} &=&\Big[(\bar\xi W_c)_{s_1}\Gamma_j h_v\Big]\,
\Big[(\bar\xi W_c)_{s_2} \frac{\not\!\!n_+}{2}\,\Gamma_{j^\prime} 
(W^\dagger_c\xi)_{s_3}\Big],
\end{eqnarray}
where $\Gamma_j=\{1,\gamma_5,\gamma_{\alpha_\perp}\}$ denotes a 
basis of the four independent Dirac matrices between 
$\bar\xi$ and $h_v$. The last two operators contain 
collinear field products at three independent light-cone positions 
$x+s_i n_+$, $i=1,2,3$. The momentum-space coefficient functions of these 
currents therefore depend on two independent variables that 
describe the distribution of the total light-cone momentum 
among the three collinear fields. These operators describe an 
effective heavy-to-light transition vertex with two additional 
transverse hard-collinear gluons ($J_{12} $), or an additional hard-collinear 
quark-antiquark pair ($J_{14}$) plus any number of $n_+\cdot A_c$ 
gluon fields from the hard-collinear Wilson line $W_c$.
The soft fields including $h_v$ are 
multipole-expanded and depend only on $x_-^\mu$. At this point, we 
can write the representation of the QCD current in SCET$_{\rm I}$ 
to any order in the $\lambda$ expansion in the form
\begin{eqnarray}
\label{matchall}
\hspace*{2cm} 
(\bar{\psi}\hspace*{0.03cm}\Gamma_i \hspace*{0.03cm}Q)(x)&=& 
e^{-i m_b v\cdot x}\,
 \sum\limits_{j,k} \widetilde{C}^{(k)}_{ij}(\hat{s}_1,
\ldots,\hat{s}_n)\otimes
J_{j}^{(k)}(\hat{s}_1,\ldots,\hat{s}_n;x),
\end{eqnarray}
\end{widetext}
\noindent
where the $\otimes$ stands for a convolution over 
a set of dimensionless variables $\hat{s}_i\equiv s_i m_b$. 
The superscript $k$
refers to the scaling of the current operator with $\lambda$ relative 
to the leading-power currents. 
The subscript $j$ enumerates the effective currents at a given order 
in $\lambda$. The two-point correlator (\ref{eq:correlator}) of QCD currents 
is accordingly written as  
\begin{equation}
T^{\mu\nu}=\tilde H_{jj'}(\hat s_1,\dots, \hat s_n)\otimes 
T^{{\rm eff},\mu\nu}_{jj'}(\hat s_1,\dots, \hat s_n).
\end{equation}
The hard function  $\tilde H_{jj'}$ is a product of 
SCET Wilson coefficients
$\tilde C_{ij}^{(k)} \tilde C_{i'j'}^{(k')}$ and 
$T^{{\rm eff},\mu\nu}_{jj'}$ is a correlator of two of the above 
SCET$_{\rm I}$ currents.

The effective correlators are computed with the SCET$_{\rm I}$ 
Lagrangian including the 
power-suppressed interactions up to $\mathcal{O}(\lambda^2)$, which 
have been derived in \cite{Beneke:2002ph,Beneke:2002ni}. The power-suppressed 
SCET$_{\rm I}$ Lagrangian terms are dealt with in the interaction picture. 
Writing 
\begin{equation} 
{\cal L}_{{\rm SCET}_{\rm I}} = {\cal L}^{(0)} + 
{\cal L}^{(1)}+{\cal L}^{(2)}+\ldots,
\end{equation}
the possible time-ordered products  
that build up the hadronic tensor  at $\mathcal{O}(\lambda^2)$
are   
\begin{eqnarray}
\label{eq:combos}
a) && J^{(0)} J^{(2)}_k + \,\mbox{sym.}
\nonumber\\
b) && J^{(1)}_k J^{(1)}_l 
\nonumber\\
c) && J^{(0)} J^{(1)}_k {\cal L}^{(1)} + \,\mbox{sym.}
\nonumber\\
d) && J^{(0)} J^{(0)} {\cal L}^{(2)} 
\nonumber\\
e) && J^{(0)} J^{(0)} {\cal L}^{(1)}  {\cal L}^{(1)}.
\end{eqnarray}
The $\mathcal{O}(\lambda)$ suppressed interactions in the effective 
Lagrangian involving hard-collinear 
($\xi$) and soft ($q_s$) quarks read 
\begin{eqnarray}
\label{eq:lagrangian1}
{\cal L}_\xi^{(1)}&=&\bar{\xi}x_{\perp}^\mu n_-^\nu W_c \,g F^s_{\mu\nu} W_c^\dagger
\frac{\not\!\!n_+}{2}\xi,\\
{\cal L}^{(1)}_{\xi q}&=&\bar q_s W_c^\dagger i\Slash{D}_{\perp c}\xi - 
\bar \xi i\ov{\Slash D}_{\perp c} W_c q_s. 
\label{eq:Lqxi}
\end{eqnarray}
At second order there are several terms. Examples are   
\begin{equation}
 {\cal L}^{(2)}_{2\xi } = \frac12\bar{\xi}x_\perp^\mu x_{\perp\rho}n_-^\nu
W_c[D_{\perp s}^{\rho},g
F_{\mu\nu}^s]W_c^\dagger\frac{\not\!\!n_+}{2}\xi,
\label{eq:Lqxi2}
\end{equation}
and 
\begin{equation}
\label{lhqet}
{\cal L}^{(2)}_{\rm HQET, mag}=\frac{C_{\rm mag}}{4 m_b}\,
\bar h_v\sigma_{\mu\nu}g F_s^{\mu\nu}h_v,
\end{equation}
where $C_{\rm mag}\not= 1$ 
represents the renormalization of the chromomagnetic 
interaction by hard quantum fluctuations. The complete list of relevant 
terms in given in~\cite{Beneke:2004in}. 

It is worth noting that the interaction ${\cal L}^{(1)}_{\xi q}$ in 
(\ref{eq:Lqxi}) describes the process $q_c \to g_c+q_s$, that is, the 
radiation of a soft quark from an energetic quark, which is thereby 
seen to be power-suppressed relative to the emission of soft 
gluons, $q_c \to q_c+g_s$. ${\cal L}^{(1)}_{\xi q}$, despite being 
power-suppressed, is a crucial term for hard-spectator scattering 
at {\em leading} 
order in the $1/m_b$ expansion discussed in previous sections, 
since it turns the soft 
spectator quark in the $B$ meson into the energetic quark that forms 
the constituent of an energetic light meson. ${\cal L}^{(1)}_{\xi}$,  
on the other hand, 
describes the correction to the soft-gluon emission 
process $q_c \to q_c+g_s$, and hence the leading correction to the 
usual soft (``eikonal'') approximation. Due to the vanishing of 
odd powers in the $\lambda$ expansion, even the second-order corrections 
in $\lambda$ are needed to obtain the $\Lambda/m_b$ term to the 
semi-leptonic decay rate, which includes a time-ordered product with two 
insertions of ${\cal L}^{(1)}$ in (\ref{eq:combos}).

By construction the time-ordered products (\ref{eq:combos}) are 
evaluated with the leading-order Lagrangian ${\cal L}^{(0)}$.
We recall that when the hard-collinear fields are redefined 
according to \cite{Bauer:2001yt}
\begin{equation}
\label{eq:redef}
\xi = Y\xi^{(0)},\quad A_c =  
Y A_c^{(0)} Y^\dagger, \quad W_c =  Y W^{(0)}_c Y^\dagger, 
\end{equation}
where $Y$ is the soft Wilson line of $n_-\cdot A_s$, the soft and 
collinear fields are decoupled in ${\cal L}^{(0)}$.
The $\bar B$ meson by definition 
contains no collinear degrees of freedom, meaning that 
the $\bar B$-meson state is represented as the tensor product 
$|\bar B\rangle\otimes |0\rangle$, where the first (second) factor 
refers to the soft (hard-collinear) Hilbert space. It follows that 
the matrix element of any SCET$_{\rm I}$ current correlator 
$T^{\rm eff,\mu\nu}$, including in general 
time-ordered products with sub-leading interactions from the 
Lagrangian, can be written in the factored form 
\begin{widetext}
\begin{eqnarray}
&&\langle\bar B| T^{\rm eff}(\hat s_1,\dots,\hat s_n)|\bar B\rangle=
i\int d^4 x d^4 y\dots e^{i(m_b v-q) x}\,\langle\bar{B}
|\bar{h}_v[{\rm soft\, fields}]h_v|\bar{B}\rangle(x_-,y_-,...)
\nonumber\\
&& \hspace*{2cm}\times\, 
\langle 0 |[{\rm hard-collinear\, fields}] |0\rangle(\hat s_1,\dots,\hat s_n;
x,y,...). 
\label{eq:csfact}
\end{eqnarray}
\end{widetext}
\noindent
The additional integrals over $d^4 y\dots$ are related to 
insertions of the power-suppressed Lagrangian terms. The soft matrix element 
depends only on $x_+\equiv n_+ x, y_+\equiv n_+ y,\ldots$, so the integrations 
over transverse positions and the $n_-$ components can be lumped 
into the definition of the collinear factor. The soft and 
collinear matrix elements are then  linked by multiple convolutions 
over the light-cone variables $x_+, y_+,\ldots$.  
Having carried out these steps, and using momentum-space representations 
of the hard coefficients, jet-functions, and shape-functions, a generic 
term in the factorization formula is a sum of convolutions
\begin{eqnarray}
T &\!\!=\!\!& \sum H( u_1,\dots u_i)\otimes 
\J(u_1,\dots,u_i;\omega_1, \dots, \omega_n)
\nonumber\\[0.2cm]
&& \hspace*{0.45cm} \otimes \,S(\omega_1,\dots,\omega_n)
\label{eq:masfact}.
\end{eqnarray}
to any order in the $\lambda$ expansion. The convolution variables 
$u_i=\np \cdot p_i/m_b$ are the fractions of longitudinal momentum 
$\np\cdot p_i$ carried by the hard-collinear fields in 
SCET$_{\rm I}$. The convolution variable $\omega_i$ is conjugate to 
$n_+ \cdot x_i$, 
and corresponds to $n_- \cdot k_i$, where $k_i$ is the 
(outgoing) momentum of a soft field.

The jet functions are defined as the vacuum matrix elements of 
hard-collinear fields as they appear in (\ref{eq:csfact}). Since the 
hard-collinear virtuality $m_b \Lambda \gg \Lambda^2$, the jet functions 
can be computed in perturbation theory. On the other hand, the soft 
$B$-meson matrix elements in (\ref{eq:csfact}) define 
non-perturbative functions. At any given order in the $\lambda$ expansion, 
the soft functions are obtained by stripping the hard-collinear fields 
(since they define the jet functions) from the time-ordered product terms. 
Up to $\mathcal{O}(\lambda^2)$, one is left with the 
$B$-meson matrix elements
\begin{widetext}
\begin{eqnarray}
&&\langle \bar B|(\bar h_v Y)(x_-)_{a\alpha} (Y^\dagger h_v)(0)_{b\beta}|\bar B\rangle,\quad 
\langle \bar B|(\bar h_v Y)(x_-)_{a\alpha} (Y^\dagger iD_s^\mu Y)(z_-)_{cd} 
(Y^\dagger h_v)(0)_{b\beta}|\bar B\rangle,
\nonumber \\[0.2cm]
&& \langle \bar B|(\bar h_v Y)(x_-)_{a\alpha} (Y^\dagger iD_{s\perp}^\mu Y)(z_{1-})_{cd}
(Y^\dagger iD_{s\perp}^\nu Y)(z_{2-})_{ef} (Y^\dagger
h_v)(0)_{b\beta}|\bar B\rangle, 
\nonumber \\[0.2cm]
&& \langle \bar B|(\bar h_v Y)(x_-)_{a\alpha} 
 (Y^\dagger h_v)(0)_{b\beta}\,(\bar q_s Y)(z_{1-})_{c\gamma} 
 (Y^\dagger q_s)(z_{2-})_{d\delta}|\bar B\rangle,
\\[-0.2cm] \nonumber
\label{derivativebasis}
\end{eqnarray}
where colour (Latin) and spinor (Greek) indices have been made explicit. 
These matrix elements are then decomposed in colour and spin 
into scalar functions, 
which depend on one, two or three $\omega$ variables.
It follows that to order $1/m_b$, but 
to arbitrary order in $\alpha_s$ in the coefficient functions, 
the semi-inclusive 
semi-leptonic differential decay rate depends on a large number of 
multi-local shape-functions. An investigation of the structure of 
the convolution shows that at this order the full-QCD hadronic 
tensor has the factorized structure 
\begin{eqnarray}
 T &=& H \cdot \,\J(\omega)\otimes S(\omega)
\nonumber \\[0.1cm]
&& +\, \sum H(u_1,u_2)\otimes 
\J(u_1,u_2;\omega)\otimes S(\omega)
+ \sum H(u)\otimes 
\J(u;\omega_1, \omega_2)\otimes S(\omega_1,\omega_2)
\nonumber \\[0.1cm]
&& +\, \sum H \cdot \, 
\J(\omega_1, \omega_2,\omega_3)\otimes S(\omega_1,\omega_2,\omega_3) 
+\ldots 
\label{eq:masfact2},
\end{eqnarray}
\end{widetext}
\noindent
where the ellipses denote $1/m_b^2$ terms not considered here, 
and for each term the dependence on the convolution variables of the 
most complicated structure is shown. The first line represents 
the leading order in the $\Lambda/m_b$ expansion. It depends 
only on a single function of a single variable. It is evident 
that the $\Lambda/m_b$ correction is much more complicated. 
It is unlikely that 
multi-dimensional soft functions can ever be reliably determined 
from data or be computed by non-perturbative methods, so some 
amount of modelling of these functions will be required.
The renormalization-group equations of the effective theory allow 
large logarithms $\ln m_b/\Lambda$ to be summed to all orders. The 
anomalous dimensions are themselves multi-dimensional integration 
kernels. There is a considerable number of them, given the 
number of different terms, which may all mix under renormalization. 
The problem of summing logarithms to all orders at $\mathcal{O}(1/m_b)$ 
has therefore so far not been addressed.

\begin{figure}
\vskip0.2cm
\begin{center}
\includegraphics[width=7.5cm]{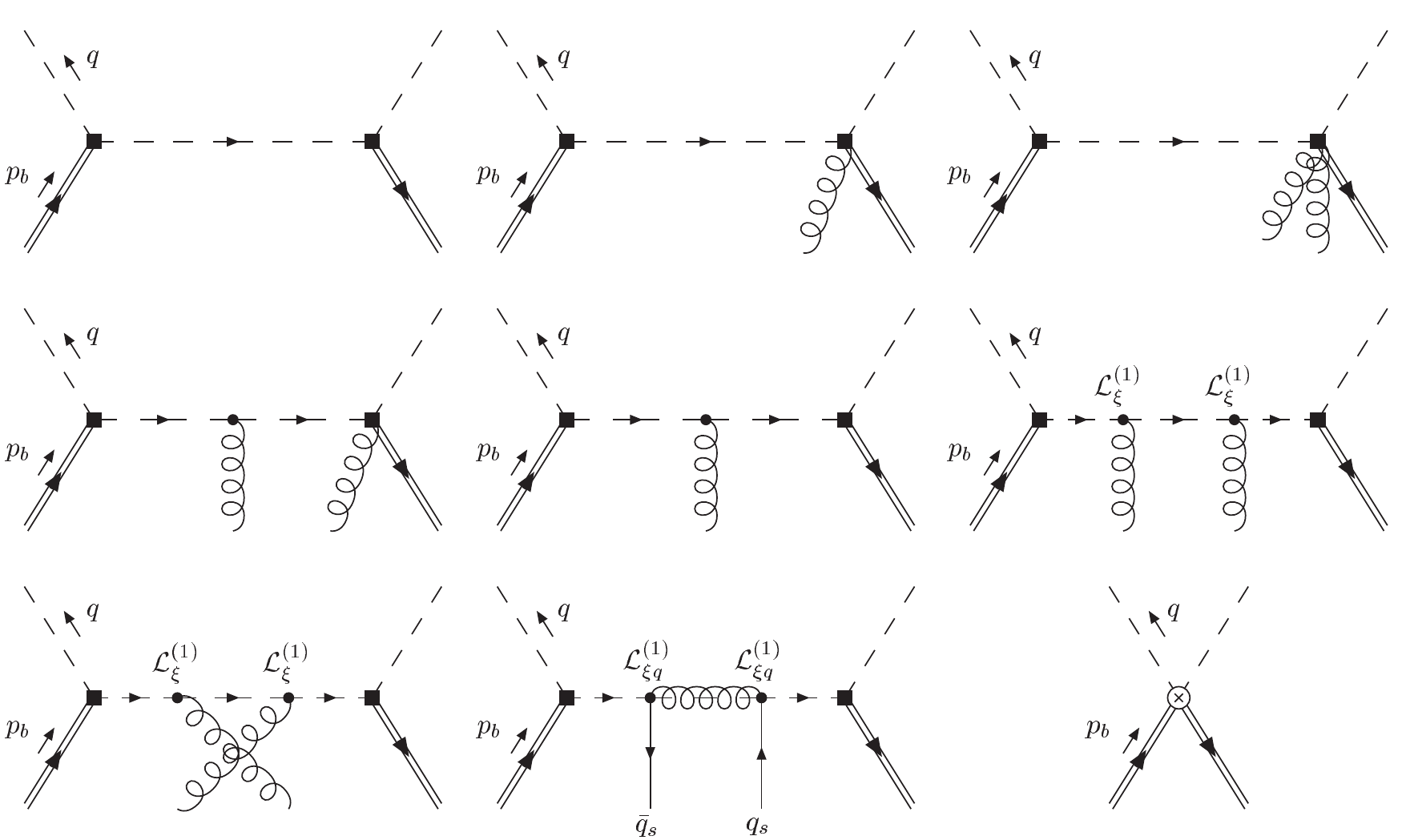}
\end{center}
\vspace*{-0.3cm}
\caption{Tree diagrams contributing to the current correlators 
$T^{{\rm eff},\mu\nu}$. 
Not shown are diagrams that vanish when $n_+ A_c=0$, $n_- A_s=0$, 
or are symmetric to those shown. External lines (except for the 
dashed current insertion) are soft. The double lines refer to the 
heavy-quark field $h_v$. Figure from \cite{Beneke:2004in}.}
\label{fig:semitree}
\end{figure}

Fortunately, the structure of the result is much simpler than 
(\ref{eq:masfact2}) in the tree approximation as was shown in 
\cite{Bosch:2004cb,Beneke:2004in}. Figure~\ref{fig:semitree} displays 
the relevant tree diagrams. The reason for this simplification is 
that tree diagrams can only have cuts with a single internal hard-collinear 
particle, and hence there are no convolutions in hard-collinear 
variables $u_i$. The convolutions of multi-dimensional soft functions 
with the tree-level jet functions can be lumped into a set of 
unknown functions of a {\em single} variable. An interesting aspect 
of this analysis is that there is a contribution from a four-quark 
HQET operator of the form $\bar h_v h_v \bar q_s q_s$ to the differential 
semi-leptonic decay rate at $\mathcal{O}(1/m_b)$ 
(see the middle diagram in the last row of 
Figure~\ref{fig:semitree}), whereas in the total decay rate such 
contributions are at least of order $1/m_b^3$. For a given flavour 
of $q_s$, these operators contribute differently to the decay of 
the charged and the neutral $B$ meson, primarily in the region of 
small $n_-\cdot P$, where $P$ is the total momentum of the hadronic 
final state $X_u$. For a numerical estimate, which requires some modelling, 
I refer to \cite{Beneke:2004in}.

Note that the operator $T^{\mu\nu}$ in (\ref{eq:correlator}) appears 
in similar form in many other processes. The SCET formalism described here 
can be adapted to address power corrections in the 
process $e^+ e^- \to $ two jets, in semi-inclusive 
deep-inelastic scattering (DIS) $e^- + p \to e^- + X$, when the hadronic 
state $X$ is collimated to form a single jet, or Drell-Yan production 
near the threshold. For DIS, the analogy is especially clear, since 
one only needs to replace the $B$-meson by a proton, and the 
weak current by the electromagnetic current. SCET provides 
the expressions for the sub-leading interactions and hard vertices 
that are needed to extend the standard factorization theorems 
based on the eikonal approximation and the factoring of jet 
functions to sub-leading power.

\section{Radiative corrections}

Significant efforts have been dedicated over the past ten years to push 
factorization calculations to $\mathcal{O}(\alpha_s^2)$ in the perturbative 
expansion. This implies two-loop calculations of the matching 
coefficient of the SCET$_{\rm I}$ current $\bar \xi h_v$ and the form-factor 
kernel $T^{\rm I}$ for hadronic two-body decays, and one-loop 
calculations of the hard- and hard-collinear matching coefficients 
in the spectator-scattering terms. In this section I review these 
efforts and summarize a few results.

\subsection{Form-factor relations}

\begin{figure}[t]
\vskip0.2cm
\begin{center}
\includegraphics[width=5.2cm]{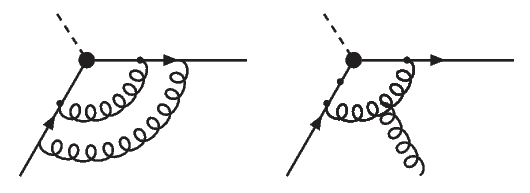}
\hskip0.2cm
\includegraphics[width=1.9cm]{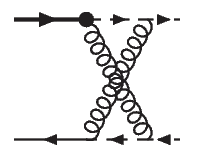}
\end{center}
\vspace*{-0.3cm}
\caption{Representative diagrams for the two-loop contribution to 
$C_i$, and the one-loop contribution to $C^{(B1)}_i$ and 
$J$ (from left to right).}
\label{fig:hardcurrentC}
\end{figure}

The matching coefficients relevant to the factorization of the 
heavy-to-light form factors can be read off 
from (\ref{ff2}), (\ref{b1match}), (\ref{eq:Tform}). 
At $\mathcal{O}(\alpha_s^2)$ one 
needs the one-loop corrections to $C^{(B1)}_i$, the coefficient functions 
of the $\bar \xi_c A_{\perp hc} h_v$ operators, and to $J$, 
defined by the non-local operator matching equation (\ref{matchrel}). 
The two pieces were obtained in \cite{Beneke:2004rc,Becher:2004kk},  
and \cite{Hill:2004if,Beneke:2005gs}, respectively. In addition, one needs 
the two-loop coefficient functions $C_i$ of the $\xi_c h_v$ operators,  
which have been calculated in 
\cite{Asatrian:2008uk,Bonciani:2008wf,Beneke:2008ei,Bell:2008ws} 
for the vector and axial-vector QCD current and in 
\cite{Bell:2010mg} for the remaining (pseudo) scalar and tensor 
currents. Since all quantities involved have operator definitions, 
these are fairly standard loop calculations with special vertices 
derived from the form of the operators and the SCET Lagrangians. 
Figure~\ref{fig:hardcurrentC} shows a representative diagram for 
each of the three calculations.

With these results the symmetry-breaking corrections to the heavy-to-light 
form factor relations \cite{Charles:1998dr} that were 
first computed at $\mathcal{O}(\alpha_s)$ in \cite{Beneke:2000wa}, can 
be extended to $\mathcal{O}(\alpha_s^2)$. This has interesting applications 
to exclusive and semi-inclusive semi-leptonic and radiative $B$ decays 
\cite{Bell:2010mg}, of which a few will be mentioned here.

Two examples of form-factor ratios, which simplify in the heavy-quark 
limit, are 
\begin{eqnarray}
{\cal R}_1(E) &\equiv& \frac{m_B}{m_B+m_P}\frac{f_T(E)}{f_+(E)},
\nonumber\\
{\cal R}_2(E) &\equiv& \frac{m_B+m_V}{m_B}\frac{T_1(E)}{V(E)}.
\label{ffratios}
\end{eqnarray}
We can write the first in the form 
\begin{equation}
{\cal R}_1(E) = R_T(E) +\int_0^1 \!d\tau \,C_{T+}^{(B1)}(\tau,E)\,
\frac{\Xi_P(\tau,E)}{f_+(E)},
\label{ffratios2}
\end{equation}
where $R_T = C_{f_T}/C_{f_+}$ is the ratio of the two-loop matching 
coefficients for the tensor and vector currents for the $B\to P$ 
(pseudo-scalar) transition, 
$C_{T+}^{(B1)}(\tau,E)=C_{f_T}^{(B1)}(\tau,E)-
C_{f_+}^{(B1)}(\tau,E)\,R_T(E)$, and $\Xi_P(\tau,E)$ is given in 
(\ref{b1match}). A similar result holds for the second 
ratio ${\cal R}_2(E)$ that applies to $B\to V$ (vector) form factors.
We recall that factorization as discussed here is valid 
when the energy of the outgoing light meson satisfies $E\gg \Lambda$. 
In this limit, the ratios above approach the value 1 plus 
perturbative corrections comprised in $R(E)$, which are independent 
of any non-perturbative parameters, and the spectator-scattering 
correction. Since this term begins at $\mathcal{O}(\alpha_s)$, 
the sensitivity to non-perturbative form factors and light-cone 
distribution amplitudes is reduced to  $\mathcal{O}(\alpha_s)$ 
in ratios. 

To be specific, consider the point of maximal recoil, i.e. $E=m_B/2$ or
$q^2=0$, at which we find
\begin{widetext}
\begin{eqnarray}
R_T(E_{\rm max}) &=& 1 + \frac{\alpha_s^{(4)}(\mu)}{4\pi}\left[\frac{8}{3} -
\frac{4}{3}L_{\nu}\right]
+\bigg(\frac{\alpha_s^{(4)}(\mu)}{4\pi}\bigg)^{\! 2} \,
\left[-\frac{100}{9}L_{\mu}L_{\nu} + \frac{200}{9}L_{\mu} + 6L_{\nu}^2 -
\frac{922}{27}L_{\nu}\, \right. \nonumber\\
&& \left. - \frac{16}{3}\zeta(3) + \frac{10}{3}\pi^4 - \frac{952}{27}\pi^2 +
\frac{8047}{162} + \frac{128}{27}\pi^2 \ln{2}\,\right]\,,
\end{eqnarray}
with $L_{\mu}=\ln(\mu^2/m_b^2)$, $L_{\nu}=\ln(\nu^2/m_b^2)$. The scale 
$\nu$ is used to distinguish the scale dependence of the QCD tensor 
current from the residual scale dependence of the truncated perturbative 
expansion. Numerically, the form factor ratio ${\cal R}_1(E_{\rm max})$ 
is (setting $\nu=\mu=m_b$) \cite{Bell:2010mg}
\begin{eqnarray}
{\cal R}_1(E_{\rm max})
&=& 1 + \Big[0.046\,(\mbox{NLO}) + 0.015\,(\mbox{NNLO})\Big]\,(R_T)\,
\nonumber\\
&& - 0.160\,\Big\{1+0.524\,(\mbox{NLO spec.})
- 0.002 \,(\delta_{\rm log}^\parallel)\Big\}
\quad = \quad 0.817.
\label{RT}
\end{eqnarray}
\end{widetext}
\noindent
In this expression we separated the symmetry-conserving term~(first
number, equal to 1), the correction to $R_{T}$ 
(remaining terms in the first line) and the spectator-scattering 
contribution (second line, including a small correction 
$\delta_{\rm log}^\parallel$ from renormalization-group 
summation~\cite{Beneke:2005gs}), and within these the 
$\mathcal{O}(\alpha_s^2)$ contributions ``NNLO'' and ``NLOspec.".

We observe that these $\mathcal{O}(\alpha_s^2)$ 
corrections are $30\%$ (first line of (\ref{RT})) and $50\%$ 
(second line) of the 
$\mathcal{O}(\alpha_s)$ ones, and result in an overall 
reduction of ${\cal R}_1(E_{\rm max})$ by 20\% relative to the 
symmetry limit. The $R_T$ and 
spectator-scattering corrections have opposite sign, but the 
latter are larger and determine the sign of the
deviation from the symmetry limit. The same observations hold for 
${\cal R}_2(E_{\rm max})$, but in this case the sign of the two terms 
is opposite and one finds ${\cal R}_2(E_{\rm max}) = 1.067$. 
The normalization of the spectator-scattering term (the number 
$-0.160$ in (\ref{RT})) involves the parameter combination
\begin{equation}
r_{\rm sp} = \frac{9 f_\pi \hat f_B}{m_b f_+^{B\pi}(0) \lambda_B},
\label{defrsp}
\end{equation}
which captures most of the dependence on non-perturbative parameters
of spectator scattering and the form-factor ratio as a whole, and 
supplies the by far dominant source of quantifiable 
theoretical uncertainty. In addition there are $\mathcal{O}(\Lambda/m_b)$ 
corrections to the form-factor factorization formula (\ref{ff2}). 

There are also QCD sum-rule calculations of the QCD form factors, which 
have different theoretical systematics, and implicitly include some 
$\mathcal{O}(\Lambda/m_b)$ power corrections. These calculations 
give~\cite{Ball:2004ye, Ball:2004rg} ${\cal
  R}_1(E_{\rm max})=0.955$ and ${\cal R}_2(E_{\rm max})=0.947$. 
For the tensor-to-vector ratio
${\cal R}_2$, one notices that the sign of the symmetry-breaking
correction is opposite for these two methods. The origin of 
this discrepancy is not understood. Before drawing conclusions 
on the sum-rule method or the size of power corrections, however, 
a dedicated analysis of form-factor ratios (rather than the form factors
themselves) with correlated theoretical uncertainties should be 
performed within the QCD sum-rule method. 

\subsection{Semi-inclusive $B\to X_s\ell^+\ell^-$ decay}

\begin{figure}[t]
\vskip0.2cm
\begin{center}
\hspace*{-0.2cm}
\includegraphics[width=7cm]{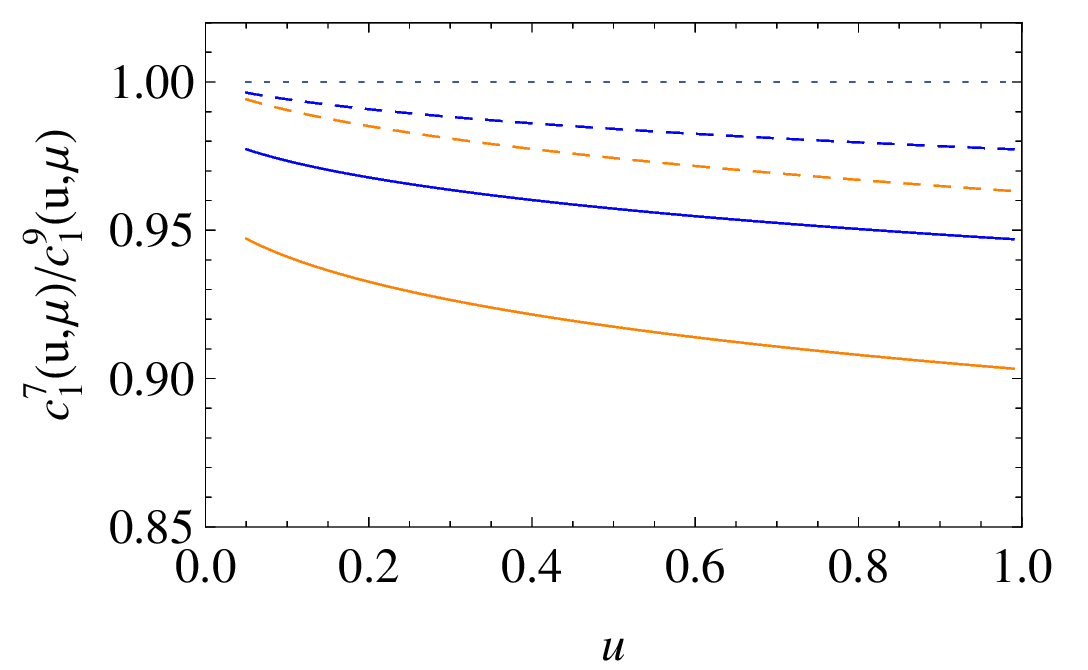}
\end{center}
\vspace*{-0.3cm}
\caption{The matching
coefficient $c^7_1(u,\mu)/c^9_1(u,\mu)$ as a function
of $u$~(related to the di-lepton invariant mass $q^2=(1-u)m_b^2$)
in the one-loop (dashed) and two-loop (solid) approximation.
The blue/dark grey curves refer to $\mu=m_b=4.8\,$GeV,
and the orange/light grey ones to $\mu=1.5\,$GeV. 
Figure from \cite{Bell:2010mg}.}
\label{fig:c7c9}
\end{figure}

The two-loop correction to the QCD current matching to SCET$_{\rm I}$ 
has an interesting application to semi-inclusive $B\to X_s\ell^+\ell^-$ 
decay. As discussed in Section~\ref{sec:semipower}, the semi-inclusive 
rate factorizes in the form 
\begin{equation}
\label{eq:factll}
\mathrm{d} \Gamma^{[0]} = h^{[0]} \times J \otimes S\,,
\end{equation}
at leading order in the $1/m_b$ expansion with hard functions 
$h^{[0]}$ that can be expressed in terms of the coefficients $C_i$ of 
$\xi_c \Gamma_i h_v$.

Here I focus on the differential 
(in $q^2$, the $\ell^+\ell^-$ invariant mass) 
forward-backward (FB) asymmetry in the angle between the
positively charged lepton and the $\bar B$ meson in the
centre-of-mass frame of the di-lepton pair in the presence of an 
upper limit on the invariant mass $m_X$ of the hadronic final state 
\cite{Lee:2005pwa,Lee:2008xc}. It is well-known that due to the interplay 
of electroweak penguin operators mediating $b\to s\ell^+\ell^-$ and 
the electromagnetic dipole operator mediating 
$b\to s \gamma (\to \ell^+\ell^-)$, 
the FB asymmetry exhibits a zero. In the presence of an invariant-mass 
cut $m_X^{\rm cut}$ the location $q_0^2$ of the zero is given 
approximately by 
\cite{Bell:2010mg}
\begin{equation}
\frac{q_0^2}{2 m_b (m_B-\langle p_X^+ \rangle)}
= - \, \frac{\mbox{Re}\,[C_7^{\rm incl}(q_0^2)]}
{\mbox{Re}\,[C_9^{\rm incl}(q_0^2)]}\,
\frac{c_1^7(u_0)}{c_1^9(u_0)}
\label{q0eq}
\end{equation}
with $u_0\equiv 1-q_0^2/(m_b(m_B-\langle p_X^+ \rangle))$ and 
$\langle p_X^+\rangle \approx p_X^{+\rm cut}/2$. This result depends on 
$m_X^{\rm cut}$ through the corresponding cut on the light-cone momentum 
component $p_X^{+\rm cut}$.

The first factor on the right-hand side of (\ref{q0eq}) is primarily 
related to 
the Wilson coefficients of the electroweak penguin and electromagnetic 
dipole operators in the weak effective Lagrangian. The second 
factor  $c^7_1(u)/c^9_1(u)$ provides a modification 
from the matching to 
SCET$_{\rm I}$ and can be expressed in terms of the coefficients $C_i$ 
known to the two-loop order. The dependence of this factor 
on the di-lepton invariant mass is shown at LO, NLO and NNLO in 
Figure~\ref{fig:c7c9} for two different renormalization scales.
The impact of the NLO correction is to shift $q_0^2$ by $-2.2$\%. 
The size of the NNLO correction is significant, in fact larger than the 
NLO correction, and shifts the zero by another $-3$\%. Including an 
estimate of power-suppressed effects, Ref.~\cite{Bell:2010mg} finds 
\begin{equation}
q_0^{\,2} = \big[(3.34 \, \ldots \, 3.40) {}^{+0.22}_{-0.25}\big]\,\mbox{GeV}^2
 \label{eq:zeroRperpNNLO}\end{equation}
for $m_X^{\rm cut} = (2.0 \ldots 1.8)\,$GeV.
This  value of the
asymmetry zero in semi-inclusive $b\to s \ell^+\ell^-$
decay is significantly smaller than for the exclusive
case (see below), where spectator scattering is responsible for a
positive shift as is the fact that in this case
$\langle p_X^+\rangle =0$ in (\ref{q0eq}).
 On the other hand the semi-inclusive zero is in the same region as
in the inclusive case~\cite{Huber:2007vv}.

\begin{figure}[t]
\vskip0.2cm
\begin{center}
\includegraphics[width=5cm]{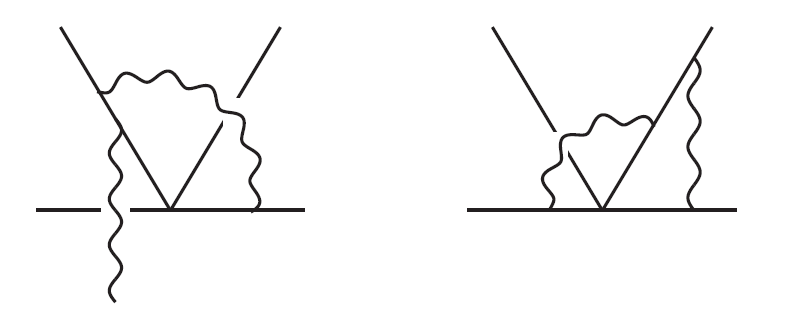}
\end{center}
\vspace*{-0.5cm}
\caption{Representative diagrams for the one-loop contribution to 
$H^{\rm II}$ (left), and the two-loop contribution to $T^{\rm I}$  
(right).}
\label{fig:charmlessdiag}
\end{figure}

\subsection{Tree-dominated charmless hadronic decays}

The matching coefficients relevant to the factorization of the 
charmless, hadronic, two-body decays can be read off 
from (\ref{fff}), (\ref{match1nonlep}), (\ref{t2}). 
At $\mathcal{O}(\alpha_s^2)$ one 
needs the one-loop correction to $H^{\rm II}$, the coefficient function 
of the $(\bar\chi\chi)( \bar \xi_c A_{\perp hc} h_v)$ 
spectator-scattering operator 
${\cal O}^{\rm II}$, and 
the two-loop correction to $T^{\rm I}$ related to the 
$(\bar\chi\chi)( \xi_c h_v)$ operator ${\cal O}^{\rm I}$. 
The hard-collinear function $J$ is the same that enters the form factor.
The computation of $H^{\rm II}$ at one-loop was performed in 
\cite{Beneke:2005vv} for the so-called tree-operators 
$Q_{1,2}$ in the weak effective Lagrangian. Two independent 
calculations can be found in \cite{Kivel:2006xc,Pilipp:2007mg}. 
The two-loop matching of $Q_{1,2}$ to the SCET$_{\rm I}$ four-quark 
operator, which defines $T^{\rm I}$, was first done for the 
imaginary part of the amplitude \cite{Bell:2007tv} and 
then in two independent calculations \cite{Bell:2009nk,Beneke:2009ek} 
for the full amplitude, finding complete agreement. These results 
together extend the 
$\mathcal{O}(\alpha_s)$ result of~\cite{Beneke:1999br,Beneke:2001ev} to 
the next order for pure tree-operator decays such as 
$B^- \to \pi^-\pi^0$. Figure~\ref{fig:charmlessdiag} shows a 
representative diagram for each of the two calculations.

The calculations are more involved than those for the form factors, 
since the presence of the second light meson adds another variable to 
the functional dependence, such that $H^{\rm II}$ depends on 
two momentum fractions, see (\ref{eq:Huv}). Furthermore, 
the four-fermion operators 
consist of two strings of fermion lines between which gluons can 
be exchanged. When factorization is done in dimensional regularization, 
one has to deal with the renormalization of operators that vanish 
in four dimensions (``evanescent operators''), whose matrix elements 
must be renormalized to zero. Related to this is the issue of consistent 
Fierz transformations in $d$ dimensions. The transition 
$b \to u(\bar u d)$, which is effected by the local operators $Q_{1,2}$ 
and where the bracket indicates the fermion fields which are 
contracted, can contribute to the $\pi^+ \pi^-$ final state in the 
form $(\bar q_s u)(\bar u d)$ (called ``right insertion'') 
as well as to the $\pi^0 \pi^0$ final state in the 
form $(\bar q_s d)(\bar u u)$ (called ``wrong insertion''), 
where $q_s = d$ is the spectator quark 
in the $\bar B_d$ meson. In the second case a Fierz rearrangement 
of the original fermion strings is required to match to the 
SCET$_{\rm I}$ operators (\ref{opsninlep1}), (\ref{opsnonlep2}). 
The simplest way to deal with this difficulty is to add the 
difference of Fierz-equivalent operators to the list of 
evanescent operators to be renormalized to 
zero \cite{Beneke:2005vv,Beneke:2009ek}. 
In the following I briefly elaborate on the issue of evanescent 
operators and the subtractions required to arrive at the renormalized 
coefficient functions. 

In case of $H^{\rm II}(u,v)$, one has to calculate the one-loop 
five-point amplitudes $b\to q_{c2}\bar q_{c2} q _{c1} g_{c1}$ 
associated with the matrix elements 
$\langle q(q_1)\bar q(q_2) q(p^\prime_1) g(p_2^\prime)|Q_{1,2}|
b(p)\rangle$, where the collinear light particles carry 
momentum $q_1=u m_b n_+ /2$, $q_2=\bar u m_b n_+/2$ in collinear-2 
direction, and $p_1^\prime =v m_b n_-/2$, 
$p_2^\prime =\bar vm_b n_-/2$ in the other. When one strips the 
SCET$_{\rm I}$ operator $O^{\rm II}(t,s)$ off all its fields,   
and represents it only by its Dirac structure, 
\begin{equation}
O_{1}\equiv 
\nmh (1 - \gamma_5) \,\otimes\, \nph(1-\gamma_5) \gp^\mu, 
\end{equation}
one finds that one-loop right-insertion amplitude also contains the operators 
\begin{widetext}
\begin{equation}
\hspace*{2cm}  O_{n} =
          \nmh \gp^\mu \gp^{\alpha_1} \ldots \gp^{\alpha_{n-1}} (1-\gamma_5)
          \,\otimes\, \nph(1-\gamma_5) \gamma_{\perp\alpha_1}
                        \ldots\gamma_{\perp\alpha_{n-1}}
\label{rightbasis}
\end{equation}
\end{widetext}
\noindent with $n=2,3,4$. All these operators are evanescent, i.e. 
vanish in four dimensions. They disappear from 
the final result, since we shall renormalize them such that 
their infrared-finite matrix elements vanish, but they must be 
kept in intermediate steps, hence the matching equation 
(\ref{match1nonlep}) has to be extended to include all four operators 
on the right-hand side. 

Evanescent operators appear already at tree level. In this
approximation the right-insertion matrix element of $Q_2$ is given by 
\vskip-0.2cm
\begin{equation}
\langle Q_2\rangle_{\rm nf}  = \frac{1}{N_c}\left(\frac{2}{\bar u}
\langle O_1\rangle - \frac{1}{u\bar u}\langle O_2\rangle
\right).
\label{treeQ}
\end{equation}
The subscript ``nf'' (for ``non-factorizable'') means 
that the ``factorizable'' diagrams with no lines 
connecting the collinear-2 and collinear-1 sectors are omitted, 
since they belong to the $T^{\rm I} O^{\rm I}$ term in (\ref{match1nonlep}) 
by virtue of the definition (\ref{opsninlep1}).
While one can simply set $\langle O_2\rangle=0$ in the above 
equation, since no $1/\epsilon$ poles are present at tree level, the 
appearance of an evanescent operator in the $d$-dimensional tree level 
matrix element implies 
that one must compute the mixing of $O_2$ into $O_1$ in the 
1-loop calculation. The renormalized coefficient function of the 
physical (non-evanescent) operator $O^{\rm II}$ can then be written as 
\begin{eqnarray} 
H^{\rm II (1)} &\!\!=\!\!& A_{1}^{(1)} - A_1^{(0)}\ast Z_{11}^{(1)} -
A_2^{(0)}\ast Z_{21}^{\rm (1)} \nonumber\\
&& + \, 2 \,T^{\rm I(1)} C_{f_+}^{(B1)(0)},
\label{fin1}
\end{eqnarray}
where $A_{n}^{(k)}$ denotes the coefficient of $O_n$ of the non-factorizable 
contribution to the ultraviolet-renormalized QCD  
$k$-loop $b\to q_{c2}\bar q_{c2} q _{c1} g_{c1}$ amplitude. 
The first term is the one-loop amplitude, which is infrared divergent. 
The next term is the SCET$_{\rm I}$ $\overline{\rm MS}$ counterterm 
$Z_{11}^{(1)}$ for the physical operator times its tree coefficient 
$A_1^{(0)}$, which subtracts this 
divergence. The third term arises from  
the mixing of the evanescent operator $O_2$ into $O_1$ at one-loop. 
The finite renormalization constant $Z_{21}^{(1)} = - 
M_{21}^{\rm (1) off}$ can be computed from the mixing matrix element 
with an infrared regularization different from the dimensional one 
(for instance, off-shell). 
Finally, the last term arises from the fact that we defined 
(\ref{opsninlep1}) to reproduce the full QCD form factor.
The stars in (\ref{fin1}) remind us that the amplitudes depend on 
momentum fractions, so the renormalization ``constants'' are actually 
convolution kernels. For the wrong insertion coefficient function, 
the expression is slightly more complicated and contains an additional 
term 
\begin{equation}
\hspace*{0.75cm} \tilde A_1^{(0)}\ast \left(\tilde M_{11}^{\rm (1) off}-
      \tilde M_{00}^{\rm (1) off}\right).
\label{fin2}
\end{equation}
It is finite and independent of the infrared regulator and ensures 
that the difference of Fierz-equivalent SCET$_{\rm I}$ operators 
is correctly renormalized to zero. Furthermore, due to the Fierz 
rearrangement there could be a contribution 
\begin{eqnarray}
&&
\tilde A_{1,\rm f}^{(1)} -A_{1,\rm f}^{(1)} + 
\tilde A_{2, \rm f}^{(0)}\ast \tilde M_{21}^{\rm (1) off}\nonumber\\ 
&& + \,\tilde A_{1, \rm f}^{(0)}\ast \left(\tilde M_{11}^{\rm (1) off}-
\tilde M_{00}^{\rm (1) off}\right),
\end{eqnarray}
from factorizable diagrams, which is not contained in the definition 
of the QCD form factor. However, this term vanishes here.

The calculation of $T^{\rm I}_i(u)$ amounts to evaluating the two-loop 
on-shell $Q_i$ matrix element of the transition $b(p)\to 
q_1(p') q_2(u q)\bar q_3(\bar u q)$ with kinematics $p=p'+q$,  
$p^2=m_b^2$, $p^{\prime \,2}=q^2=0$. The ultraviolet and 
infrared finite short-distance 
coefficient is obtained from 
\begin{eqnarray}
T_i^{(0)} &\!\!=\!\!& A^{(0)}_{i1} \, , 
\nonumber \\
T_i^{(1)} &\!\!=\!\!& A^{(1){\rm nf}}_{i1}+ Z_{ij}^{(1)} \, A^{(0)}_{j1} \, , 
\nonumber \\
T_i^{(2)} &\!\!=\!\!& A^{(2){\rm nf}}_{i1} + Z_{ij}^{(1)} \, A^{(1)}_{j1} + 
Z_{ij}^{(2)} \, A^{(0)}_{j1} + Z_{\alpha}^{(1)} \, A^{(1){\rm nf}}_{i1}
\nonumber \\ 
&&+ \, (-i) \, \delta m^{(1)} \, A^{\prime (1){\rm nf}}_{i1}
\nonumber \\ 
&& - \,T_i^{(1)} \big[C_{FF}^{(1)} +Y_{11}^{(1)}-Z_{ext}^{(1)}\big] 
\nonumber \\ &&
- \,\sum_{b>1} H_{ib}^{(1)} \, Y_{b1}^{(1)} \, , \label{eq:hsk2loopRI}
\end{eqnarray}
where now $A^{(k)}$ denotes the {\em bare} $k$-loop QCD amplitude, $Z_{ij}$, 
$\delta m^{(1)}$  the QCD renormalization matrices 
including those for the operators $Q_i$ and the bottom 
mass counterterm, 
$Y_{ab}$ the SCET$_{\rm I}$ renormalization factors, and $H^{(1)}_{ib}$ the one-loop 
coefficients of evanescent SCET$_{\rm I}$ four-quark operators. 
Again convolutions 
are implied, for instance, $H_{ib}^{(1)} \, Y_{b1}^{(1)}$ 
must be interpreted as the convolution product
$\int_0^1 du' \,H_{ib}^{(1)}(u') \, Y_{b1}^{(1)}(u',u)$. The corresponding 
equation for the two-loop wrong insertion matrix element is significantly 
more complicated, as the additional terms from Fierz rearrangement do 
not vanish any more, and the difference of two-loop SCET$_{\rm I}$ 
counterterms must be computed~\cite{Beneke:2009ek}.

The computation of the bare two-loop QCD matrix elements 
$A^{(2){\rm nf}}_{i1}$ is technically the most challenging part of 
(\ref{eq:hsk2loopRI}) and has been done by reducing tensor integrals 
to scalar integrals by Passarino-Veltman reduction~\cite{Passarino:1978jh}, 
the reduction of scalar integrals to master integrals using the Laporta 
algorithm~\cite{Laporta:2001dd} based on 
integration-by-parts~(IBP) 
identities~\cite{Tkachov:1981wb,Chetyrkin:1981qh}, and finally 
by evaluating the master integrals directly, or through Mellin-Barnes 
representations in the more complicated cases. It is convenient 
to compute the convolutions 
\begin{equation}
\int_0^1 \!\!du \; 6 u (1-u)C^{(3/2)}_n(2u-1)\, 
T_{i}^{(k)}(u)
\label{ticonv}
\end{equation}
of the kernels with the first few terms in the Gegenbauer expansion
\begin{equation}
\phi_{M}(u) = 6u \bar u \,\Big[1+\sum_{n=1}^{\infty}a_n^{M} \, 
C^{(3/2)}_n(2u-1)\Big] 
\label{gegenbauerexp}
\end{equation}
of the light-meson LCDA to express the final result in terms of the 
first two Gegenbauer moments $a_{1,2}^{M}$. In~\cite{Beneke:2009ek} 
fully analytic expressions after integration over the Gegenbauer expansion,
including the exact dependence on the charm-quark mass, have been obtained.

Putting these results together, we can proceed to the investigation of 
the so-called 
topological tree amplitudes with $\mathcal{O}(\alpha_s^2)$ accuracy. 
For definiteness, only the three $B\to \pi\pi$ decay modes will be 
considered. The decay amplitudes are given by
\begin{eqnarray}
   \sqrt2\,{\cal A}_{B^-\to\pi^-\pi^0}
   &\!\!=\!\!& A_{\pi\pi} \lambda_u\Big[\alpha_1+\alpha_2\Big], 
   \nonumber\\[0.2cm]
   {\cal A}_{\bar B^0\to\pi^+\pi^-}
   &\!\!=\!\!&A_{\pi\pi} 
       \left\{\lambda_u \Big[\alpha_1 + \hat \alpha_4^u \Big] +
         \lambda_c\,\hat \alpha_4^c\right\}, 
   \nonumber\\[0.2cm]
   -\,{\cal A}_{\bar B^0\to\pi^0\pi^0}
   &\!\!=\!\!& A_{\pi\pi}
       \left\{\lambda_u \Big[\alpha_2 
    - \hat\alpha_4^u\Big] - 
         \lambda_c\,\hat\alpha_4^c \right\}, 
\label{pirhoampsimp}
\end{eqnarray}
where $A_{\pi\pi}\equiv i \,G_F m_B^2 f_\pi f^{B\pi}_+(0)/\sqrt{2}$ 
and $\lambda_p = V_{pb} V_{pd}^*$. 
The dominant amplitudes governing the set of $\pi\pi$ amplitudes 
are the so-called colour-allowed (colour-suppressed) tree amplitude 
$\alpha_1(\pi\pi)$ ($\alpha_2(\pi\pi)$) and the non-singlet QCD 
penguin amplitudes $\alpha_4^p(\pi\pi)$. 
The equation does not show some 
smaller amplitudes that are taken into account in 
the numerical evaluation of the branching fractions below.
(Full expressions for the amplitudes are given in \cite{Beneke:2003zv}.) 
The normalization of the topological tree amplitudes is 
$\alpha_1=1$ and $\alpha_2=1/3$ in the approximation by 
tree-level $W$-boson exchange, 
when renormalization-group evolution of the $Q_i$ from the electroweak 
scale to $m_b$ is neglected. The NNLO QCD factorization results 
for the $\pi\pi$ final states read~\cite{Beneke:2009ek}
\begin{widetext}
\begin{eqnarray}
 \alpha_1(\pi\pi) &\!\!=\!\!&   1.009 + \left[ 0.023+ \, 0.010 \, i\right]_{\rm NLO}
+ \left[0.026+ \, 0.028 \, i\right]_{\rm NNLO}
\nonumber \\
&& -\, \left[\frac{r_{\rm sp}}{0.445}\right]
\left\{\left[0.014\right]_{\rm LOsp}+
\left[0.034 +\,0.027 i\,\right]_{\rm NLOsp}+
\left[0.008\right]_{\rm tw3}\right\}
= 1.000^{+0.029}_{-0.069} + (0.011^{+0.023}_{-0.050})i \, ,
\label{a2numpart1}\\[0.3cm]
\alpha_2(\pi\pi) &\!\!=\!\!&   0.220 - \left[ 0.179+ \, 0.077 \, i\right]_{\rm NLO}
- \left[0.031+ \, 0.050 \, i\right]_{\rm NNLO}
\nonumber\\
&& +\, \left[\frac{r_{\rm sp}}{0.445}\right]
\left\{\left[0.114\right]_{\rm LOsp}+
\left[0.049 +\,0.051 i\,\right]_{\rm NLOsp}+
\left[0.067\right]_{\rm tw3}\right\}
= 0.240^{+0.217}_{-0.125} + (-0.077^{+0.115}_{-0.078})i \, .
\label{a2num}
\end{eqnarray}
\end{widetext}
\noindent 
The first line of each of the two equations corresponds to the 
contribution from the operators $(\bar\chi\chi)(\bar\xi_c h_v)$ 
with coefficient function $T^{\rm I}$, while the second line accounts 
for spectator scattering at ${\cal O}(\alpha_s)$ (``LOsp'') and 
${\cal O}(\alpha_s^2)$  (``NLOsp'') and a certain power correction (``tw3'').
The numerical result sums all contributions and provides an estimate of
the theoretical uncertainties from the parameter variations
as detailed in~\cite{Beneke:2009ek}.

\begin{figure}[t]
\vskip0.2cm
\begin{center}
\includegraphics[width=7cm]{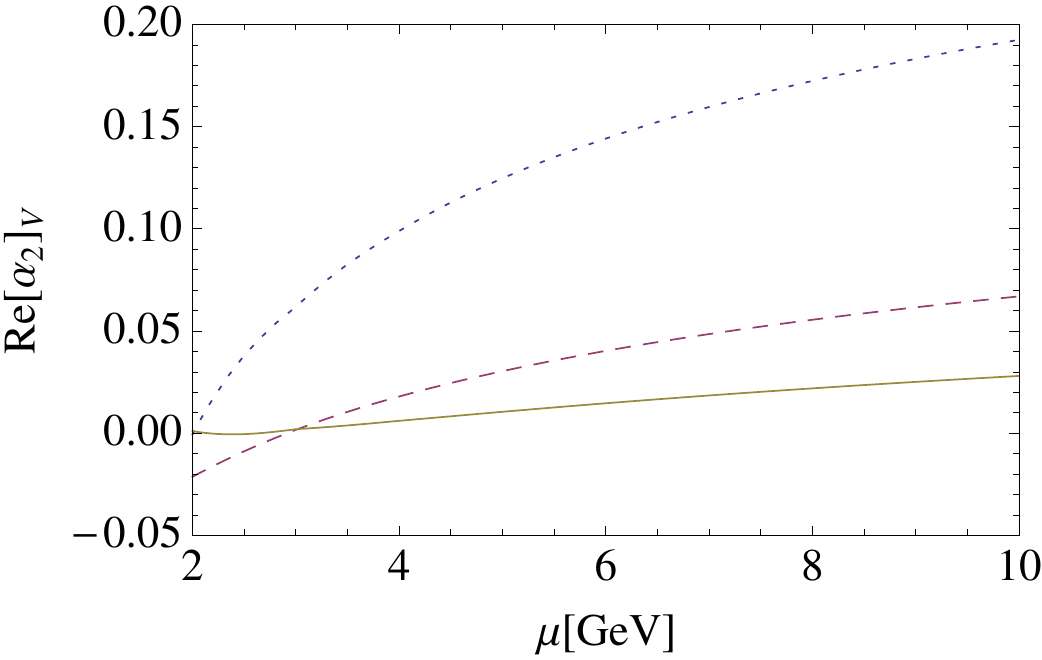}

\vspace*{0.2cm}
\includegraphics[width=7cm]{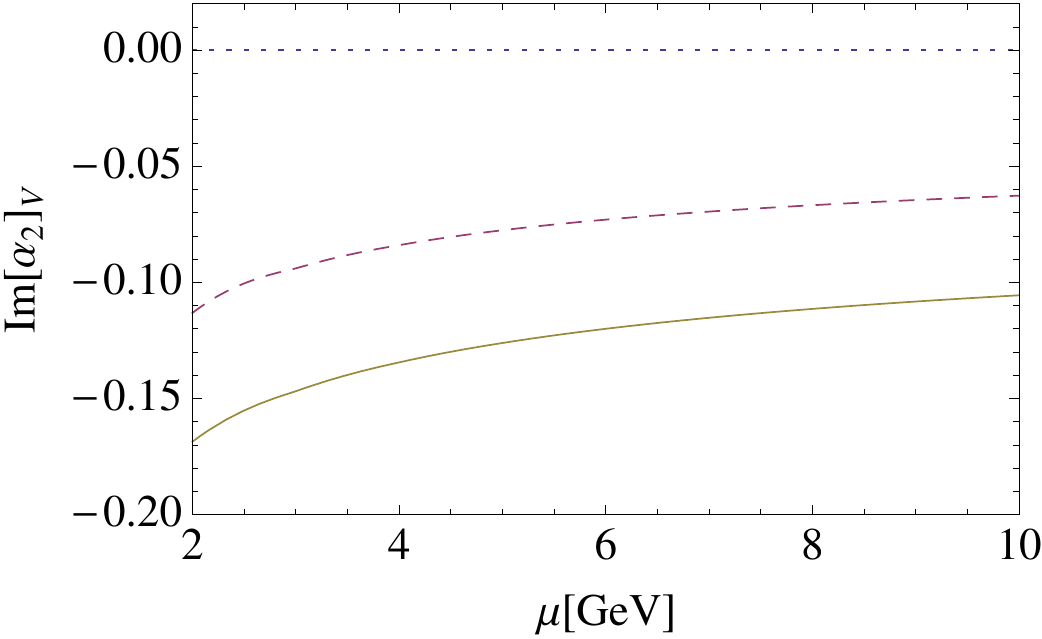}
\end{center}
\vspace*{-0.3cm}
\caption{\label{fig:scale} Dependence of the
topological tree amplitude $\alpha_2(\pi\pi)$
on the hard scale $\mu$ (vertex correction $T^{\rm I}$ only). 
The dotted, dashed and solid
lines refer to the theoretical predictions at LO, NLO and NNLO, 
respectively. Figure from~\cite{Beneke:2009ek}.}
\end{figure}

The amplitudes $\alpha_{1,2}$ are scale- and scheme-independent 
physical quantities. 
Figure~\ref{fig:scale} shows the residual renormalization scale dependence 
at LO, NLO and NNLO of the vertex correction $T^{\rm I}$ to 
the colour-suppressed amplitude $\alpha_2$. The NLO contributions 
to this amplitude are large, because they are proportional to the 
tree operator with large Wilson coefficient in the weak effective 
Lagrangian. The scale dependence stabilizes for the real part at 
NNLO, while the absorptive part receives another sizable correction, 
approximately $25\%$ of the leading-order real part, and hence its 
scale-dependence is barely reduced.
By comparing the theoretical uncertainty of the full expressions 
in (\ref{a2numpart1}), (\ref{a2num}) above 
to the one from the vertex correction alone, see Figure~\ref{fig:scale},
we see that it arises primarily from spectator scattering. The 
main contributors to the uncertainty are the parameter combination 
(\ref{defrsp}), 
which appears as an overall normalization 
factor of the spectator-scattering term, 
the second Gegenbauer moment $a^{\pi}_2(2\,\mbox{GeV})$, 
and the ``tw3'' estimate the power correction.

We observe that the vertex and spectator-scattering corrections come 
with opposite sign, both at NLO and NNLO, and separately for the 
real and imaginary part. This seems to be a general pattern that was 
also observed for the form factor ratios discussed in the previous 
subsection. Here it is 
somewhat unfortunate, since an enhancement rather than a cancellation
in the colour-suppressed tree amplitude would have been welcome
in view of the trend indicated by experimental data, as will be seen
below. 

The colour-suppressed tree amplitude $\alpha_2(\pi\pi)$ in particular 
exhibits an interesting structure. It starts out with a positive
real value 0.220 at LO. After adding the perturbative corrections to the
four-quark vertex (the first line of (\ref{a2num})), it is
found to be almost purely imaginary, $0.01 - 0.13 i$. Then 
the spectator-scattering mechanism regenerates a real part of
roughly the original size and cancels part of the strong phase.
The net result of this is that the colour-suppressed tree amplitude
can become sizable in QCD factorization when $r_{\rm sp}$ is
large, but since this enhances the cancellation of the imaginary
part, one cannot have both, a large magnitude and a large strong phase.
In essence, the dynamics of the colour-suppressed tree amplitude is 
completely governed by quantum effects, and the 
theoretical uncertainty is correspondingly large.
In comparison, the colour-allowed tree amplitude $\alpha_1(\pi\pi)$
is rather stable against radiative corrections, and never
deviates by a large amount from its LO estimate.

\tabcolsep0.35cm
\begin{table}
  \centering
  \let\oldarraystretch=\arraystretch
  \renewcommand*{\arraystretch}{1.3}
{\small \begin{tabular}{lll}
\hline
        & \phantom{0}Theory
        & Experiment
\\
\hline 
$B^-\to\pi^-\pi^0$
 & $\phantom{0}5.82_{\,-0.06\,-1.35}^{\,+0.07\,+1.42}\;\;(\star)$
 & $\phantom{0}5.48^{+0.35}_{-0.34}$ \\
$\bar B_d^0\to\pi^+\pi^-$
 & $\phantom{0}5.70_{\,-0.55\,-0.97}^{\,+0.70\,+1.16}\;\;(\star)$
 & $\phantom{0}5.10 \pm 0.19$ \\
$\bar B_d^0\to\pi^0\pi^0$
 & $\phantom{0}0.63_{\,-0.10\,-0.42}^{\,+0.12\,+0.64}$
 & $\phantom{0}1.17 \pm 0.13$
       \\
\hline
\end{tabular}}
\let\arraystretch=\oldarraystretch
\caption{
CP-averaged branching fractions (BrAv) in units of
$10^{-6}$ of $B\to \pi\pi$ decays. 
The first error on a quantity comes from the CKM parameters, while the 
second one stems from all other parameters added in 
quadrature.  
Theory corresponds to Theory II in~\cite{Beneke:2009ek} 
and adopts the values $f^{B\pi}_+(0)=0.23\pm 0.03$, 
$\lambda_B(1\,\mbox{GeV})=(0.20^{+0.05}_{-0.00})\,$GeV 
for the pion form factor and $B$-meson LCDA parameter, 
respectively. In order to focus on the hadronic uncertainty, 
the ranges given refer to the uncertainty of 
$\mbox{BrAv}(\bar B \to f)/|V_{ub}|^2$. Ranges marked by 
an asterisk $(\star)$ handle the 
dependence on the heavy-to-light form factors in a similar way and 
refer to  $\mbox{BrAv}(\bar B \to f)/(|V_{ub}|^2f_+^{B\pi}(0)^2)$.}
\label{tab:results}
\end{table}

Strictly speaking, only the charged $B$-meson tree-operator decays 
to pions and longitudinally polarized rho mesons can presently be 
predicted with complete NNLO accuracy, since the penguin amplitude 
$\alpha_4^p$ that appears in the other two modes of 
(\ref{pirhoampsimp}) is not yet completely known at 
$\mathcal{O}(\alpha_s^2)$ (see next subsection). By using the 
best available approximations for $\alpha_4^p$ the full set of 
$\pi\pi$, $\pi\rho_L$ and $\rho_L\rho_L$ final states has been 
analyzed in \cite{Beneke:2009ek,Bell:2009fm}. Table~\ref{tab:results}
shows an excerpt of the predicted CP-averaged branching fractions  
compared to present measurements compiled from BABAR and BELLE 
(including CDF and LHCb for $\pi^+\pi^-$) data as given 
in~\cite{Amhis:2014hma}. Within uncertainties the agreement between 
theory and measurement is very good for the final states with a 
substantial contribution from $\alpha_1$. The colour-suppressed 
amplitude $\alpha_2$ appears to be predicted too small, although 
the comparison recently improved due to a (preliminary) 
revision of the previous BELLE measurement. The uncertainties in 
Table~\ref{tab:results} are strongly correlated among the three decay 
modes. Further information can be 
brought to light by considering ratios of branching fractions 
sensitive to certain parameters and amplitudes. I refer to 
\cite{Beneke:2009ek,Bell:2009fm} for the details of this analysis 
and results for final states containing rho mesons.

\subsection{Penguin-dominated charmless hadronic decays}

Direct CP asymmetries require the interference of amplitudes with 
different CKM factors, $\lambda_u$ and $\lambda_c$ in 
(\ref{pirhoampsimp}), as well as a phase difference of the 
corresponding hadronic matrix elements. This necessitates the 
computation of the matrix elements of the so-called penguin operators 
$Q_{3-10}$ and dipole operators $Q_{7\gamma}$, $Q_{8g}$ in the 
weak effective Lagrangian.

At $\mathcal{O}(\alpha_s^2)$ the one-loop coefficients $H^{\rm II}_i$ 
and two-loop coefficients $T^{\rm I}_i$ for $Q_{3-10}$ (one loop less 
for $Q_{7\gamma}$ and $Q_{8g}$) are needed. They involve, in addition 
to diagram topologies identical to those relevant to $Q_{1,2}$ such as 
shown in Figure~\ref{fig:charmlessdiag}, the ``penguin contractions'' 
of same-flavour $q\bar q$ fields from the operators $Q_i$. On the 
one hand, the penguin contractions are topologically simpler than 
the vertex diagrams, on the other hand the presence of an internal massive 
charm or bottom quark loop introduces another dimensionless ratio of 
scales, which complicates the analytic calculation of the master 
integrals. The calculation of the hard spectator-scattering kernels  
for the non-singlet QCD and the colour-allowed and colour-suppressed 
electroweak penguin amplitudes has been performed in \cite{Beneke:2006mk}, 
but the two-loop calculation of $T^{\rm I}_i$ has not yet been 
completed. Technical results on the relevant master integrals have 
already appeared \cite{Bell:2014zya}. Thus, at present, the 
$\mathcal{O}(\alpha_s)$ results 
of~\cite{Beneke:1999br,Beneke:2001ev,Beneke:2003zv} for the penguin 
amplitudes and direct CP violation cannot be extended to the next order.

I therefore refrain from a detailed discussion of the numerical 
impact of the spectator-scattering contribution (see \cite{Beneke:2006mk}) 
except for one remark. In principle, the one-loop correction to 
$\alpha_4^c$ can be expected to be rather large, since there is 
a contribution from the tree operator $Q_1$ with coefficient $C_1\sim 1$ 
ten times larger than the coefficients $C_{3-6}$ of the QCD penguin 
operators. However, there is an almost complete, possibly accidental 
cancellation between the contributions from $Q_1$ with colour factors 
$C_F=4/3$ and $C_A=3$, respectively.

\section{Phenomenology}

Since the focus of this article is on the development of factorization 
and radiative corrections, and since the phenomenology of large sets 
of final states is not yet possible at NNLO in the absence of the 
calculation of at least the QCD penguin amplitude with this accuracy, 
this section will be brief, drawing in part on a summary prepared for 
\cite{Bevan:2014iga}, highlighting what in my judgement are 
interesting conclusions from and applications of $\mathcal{O}(\alpha_s)$ 
(NLO) calculations.

\subsection{Charmless, hadronic two-body decays}

Charmless, hadronic two-body decays provide by far the most observables 
due to the sheer number of possible final states. Two-body here includes 
final states with unstable particles like kaons, rho mesons etc. as long as 
they can be clearly identified as sharp resonances. 
NLO QCD factorization results are available for a variety of complete 
sets of final states: two pseudo-scalar mesons (PP) and pseudo-scalar 
plus one vector meson (PV) \cite{Beneke:2003zv}, two vector mesons (VV) 
\cite{Beneke:2006hg}, final states with a scalar meson \cite{Cheng:2007st}, 
an axial-vector meson \cite{Cheng:2007mx,Cheng:2008gxa}, and a 
tensor meson \cite{Cheng:2010yd}. 

A key issue for phenomenology is the treatment of power corrections, since  
factorization as embodied by (\ref{fff}) is not 
expected to hold at sub-leading order in $1/m_b$. 
Some power corrections related to scalar currents, which appear after 
Fierz rearrangements, are enhanced by large (``chirally enhanced'') 
factors such as 
$m_\pi^2/((m_u+m_d)\Lambda)$. Some corrections 
of this type, in particular those related to scalar penguin 
amplitudes nevertheless appear to be calculable and turn out to be
important numerically. On the other hand, 
attempts to compute sub-leading 
power corrections to hard spectator-scattering in perturbation theory 
usually result in infrared divergences, which signal the breakdown 
of factorization. These effects are then estimated and included into the 
error budget, see the contributions marked ``tw3'' in (\ref{a2numpart1}), 
(\ref{a2num}). All weak 
annihilation contributions belong to this class of effects and often
constitute the dominant source of theoretical error, in particular for
the direct CP asymmetries.
Factorization as above applies to pseudo-scalar flavour-non-singlet 
final states and to the longitudinal polarization amplitudes for vector 
mesons. Final states with $\eta$, $\eta^\prime$ require additional 
considerations, but can be included~\cite{Beneke:2002jn}. The 
transverse helicity amplitudes for vector mesons are formally
power-suppressed but can be sizable~\cite{Kagan:2004uw}, and they do not 
factorize in a simple form~\cite{Beneke:2006hg}. The description of
polarization is therefore more model-dependent than branching
fractions and CP asymmetries. Besides these conceptual uncertainties,
the lack of precise knowledge of quantities such as $|V_{ub}|$, 
heavy-to-light form factors, and the $B$-meson LCDA parameter 
$\lambda_B$ cause a significant theoretical uncertainty.

Much of the phenomenology of the most widely considered PP, PV and VV  
final states is determined by the non-singlet {\em QCD penguin amplitude} 
$\alpha_4^p$. This amplitude is certainly underestimated 
at NLO in $\alpha_s$ and leading order in the heavy-quark expansion. 
The power-suppressed but chirally-enhanced 
scalar penguin amplitude, and probably a (difficult to disentangle) 
weak annihilation contribution is required to explain the 
penguin-dominated PP final states. While the scalar penguin amplitude 
is calculable, some uncertainty remains. An important observation 
is the smaller size of the PV, VP and VV penguin amplitudes 
as compared to PP final states, which can be inferred from the
measured branching fractions of hadronic $b\to s$ transitions. This is a clear
indication of the relevance of factorization, which predicts 
this pattern as a consequence of the quantum numbers of the operators 
$Q_i$. If the penguin amplitude were entirely non-perturbative, 
no pattern of this form would be expected. A similar statement 
applies to the $\eta^{(\prime)} K^{(*)}$ final states, where factorization
explains naturally the strikingly large differences in branching 
fractions, including the large $\eta^\prime K$ branching fraction, 
through the interference of penguin amplitudes, although sizable 
theoretical uncertainties remain~\cite{Beneke:2002jn}. 
The flavour-singlet penguin amplitude seems to play a sub-ordinate 
role in these decays.

The situation is much less clear for the 
{\it strong phases and direct CP asymmetries}. A generic qualitative
prediction is that the strong phases are small, since they arise 
through either loop effects ($\alpha_s(m_b)$) or power corrections 
($\Lambda/m_b$). Enhancements may arise, when the leading-order 
term is suppressed, for instance by small Wilson coefficients. 
This pattern is indeed observed. Quantitative predictions 
have met only partial success. The observed direct CP asymmetry 
in the $\pi^+\pi^-$ and the asymmetry difference in the 
$\pi^0 K^+$, $\pi^- K^+$ final states are prominently larger than predicted. 
A comparison of all CP asymmetry results shows a presently 
ununderstood pattern of quantitative agreements and disagreements. 
Since $\alpha_s(m_b)/\pi$ and $\Lambda/m_b$ are roughly of the same 
order, it is quite possible that power corrections are ${\cal O}(1)$ 
effects relative to the perturbative calculation, preventing a
reliable quantitative estimate. However, 
the  direct CP asymmetry calculations 
are still LO calculations, contrary to the branching fractions, 
so the final verdict must await the completion of the 
NLO asymmetry calculation, which requires the two-loop computation of 
$\alpha_4^p$. Contrary to direct CP asymmetries, 
the $S$ parameter that appears in time-dependent CP asymmetries 
is predicted more reliably, since it does not require the 
computation of a strong phase. This is exploited successfully in 
NLO computations of the difference between $\sin\,2\beta$ from 
$b\to s$ penguin-dominated and from $b\to c\bar c s$ tree decays 
\cite{Beneke:2005pu,Cheng:2005bg}, and the direct determination of 
the CKM angle $\gamma$ ($\alpha$) from time-dependent CP 
asymmetry measurements for the $\pi^+\pi^-$, $\pi^\pm\rho^\mp$ 
and $\rho_L\rho_L$ final states (\cite{Beneke:2003zv} and 
Figure~1 of \cite{Beneke:2007zz}).

{\it Polarization} in $B\to VV$ decays was expected 
to be predominantly longitudinal, since the transverse helicity 
amplitudes are $\Lambda/m_b$ suppressed due to the V-A structure of 
the weak interaction and helicity conservation in short-distance 
QCD. While this is parametrically true (with one exception, see below), 
closer inspection shows that 
the parametric suppression is hardly realized in practice 
for the penguin amplitudes~\cite{Kagan:2004uw}. This leads to the qualitative 
prediction (or rather, in this case, postdiction) that 
the longitudinal polarization fraction should be close to 1 
in tree-dominated decays, but can be much less, even less than 
0.5, in penguin-dominated decays, as is indeed observed. 
However, quantitative predictions of polarization fractions 
for penguin-dominated decays must be taken with a grain of salt, 
since they rely on model-dependent or universality-inspired 
assumptions of the non-factorizing transverse helicity 
amplitudes~\cite{Beneke:2006hg}.

There is an exception to the power counting for the helicity amplitudes, 
which arises from a contribution of the electromagnetic dipole operator 
$Q_{7\gamma}$ to the transverse electroweak penguin amplitude 
$\alpha_{3,\rm EW}^{p\mp}$~\cite{Beneke:2005we}, which is {\em enhanced} 
by a factor $m_b/\Lambda$ relative to the longitudinal amplitude 
(but proportional to the small electromagnetic coupling). 
This effect manifests itself in branching and polarization fractions 
of the $\rho K^*$ final states and should be seen by comparing 
precise measurements to theoretical predictions and measurements of 
the related 
PP, PV modes $\pi K$, $\rho K$, $\pi K^*$. In particular, polarization 
measurements would then result in an indirect measurement of 
the (virtual) photon's polarization in the electromagnetic dipole transition, 
and hence be sensitive to the possible presence of a dipole 
operator with opposite chirality of the light quark due to non-standard 
physics.

\subsection{$B\to \gamma \ell\nu$}

\begin{figure}[t]
\vskip0.2cm
\begin{center}
\includegraphics[width=7cm]{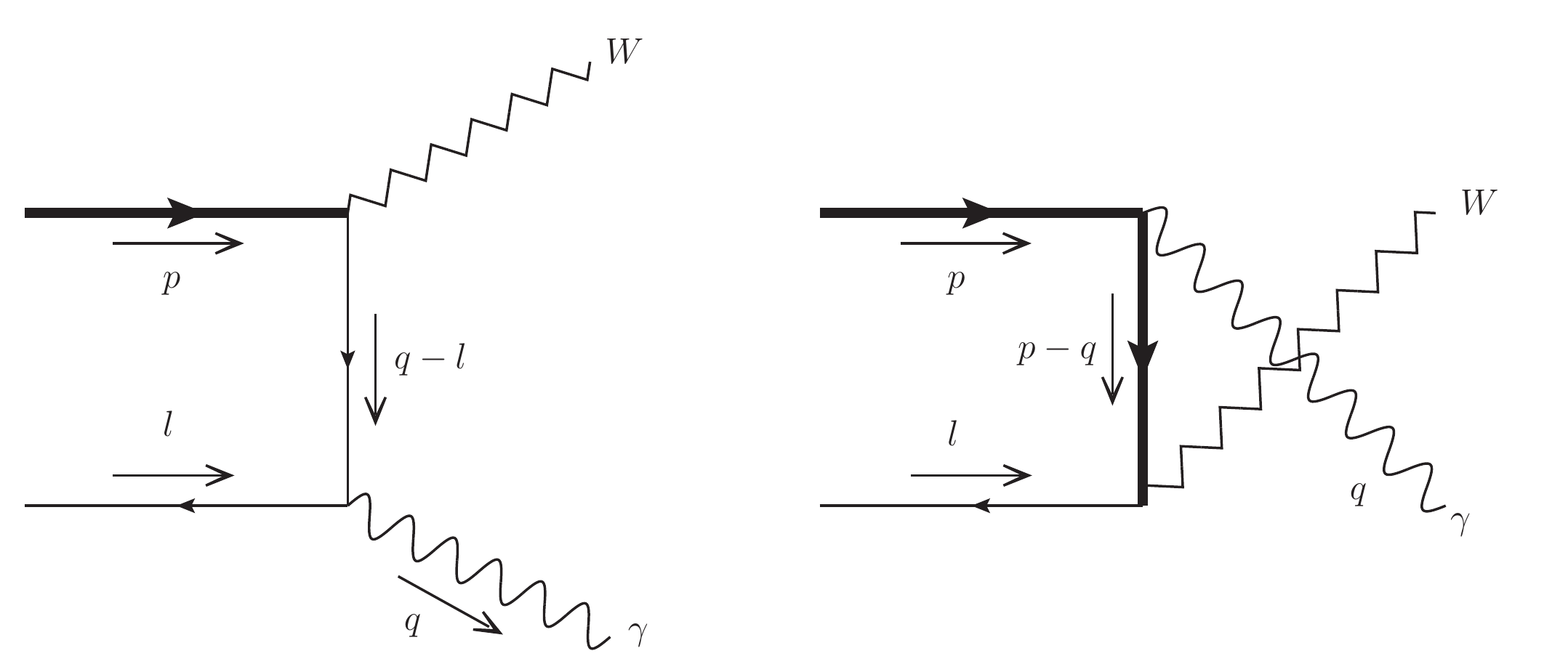}
\end{center}
\vspace*{-0.3cm}
\caption{Leading-order diagrams for the $B\to \gamma W^* (\to \ell\nu)$ 
transition. The left graph shows the 
leading-power contribution from photon emission from the up   
antiquark. Emission from the heavy $b$-quark (right) is  
power-suppressed. Figure from~\cite{Beneke:2011nf}.} 
\label{fig:LOBgamma} 
\end{figure}

The decay $B\to \gamma \ell\nu$ is accessible to factorization methods 
when the energy $E_\gamma$ of the photon is large compared to $\Lambda$. 
Even though the final state does not contain a hadron, the coupling of 
the on-shell photon to soft and collinear quarks is non-perturbative 
and leads to hadronic $\langle 0|\ldots|\bar B\rangle$ 
matrix elements of non-local operators. The two tree diagrams relevant 
to the process are shown in Figure~\ref{fig:LOBgamma}. Photon emission 
from the heavy quark (right diagram) is power-suppressed relative 
to the emission from 
the light quark in the left diagram. The differential branching fraction 
can be expressed in complete generality in terms of two independent 
$B\to \gamma$ form factors $F_{V,A}(E_\gamma)$.

Soft-collinear factorization 
properties of the $B\to \gamma \ell\nu$ amplitude have first been 
discussed in \cite{Korchemsky:1999qb}, and in 
\cite{Lunghi:2002ju,Bosch:2003fc} with SCET methods. At leading order 
in the $1/m_b$ expansion, the two form factors coincide and satisfy 
a factorization formula
\begin{equation} 
F = C \cdot J\otimes \phi_B.
\end{equation}
In the tree diagram in the left Figure~\ref{fig:LOBgamma} the hard function 
$C$ is the weak-decay vertex, while the jet function contains the 
hard-collinear light-quark propagator that connects the $W$-boson 
and photon vertices. As in other applications of leading-power factorization, 
the $B$-meson LCDA enters only through $\lambda_B$ and the logarithmic 
moments (\ref{BLCDAmom}).

\begin{figure}[t]
\vskip0.2cm
\begin{center}
\includegraphics[width=7cm]{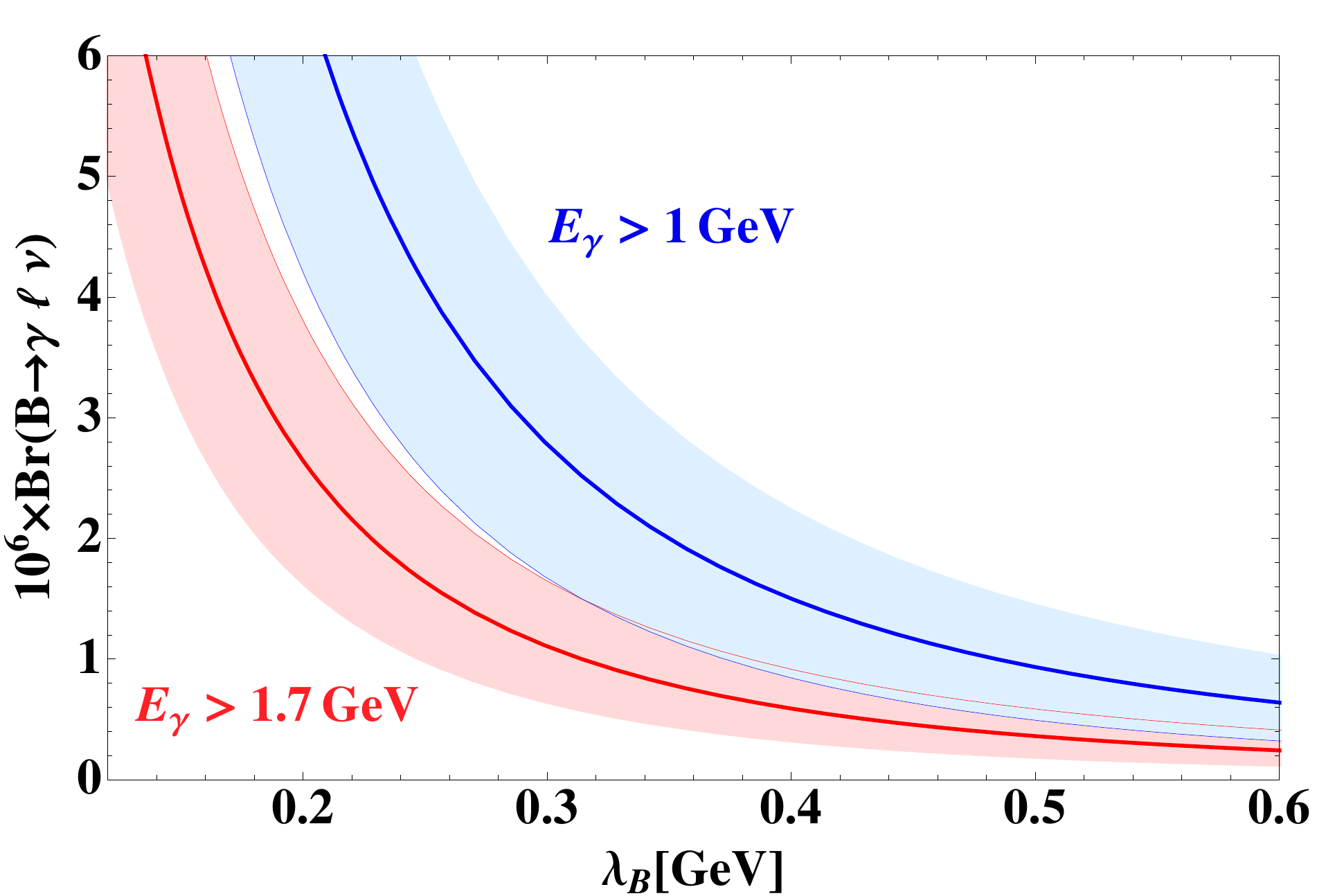}
\end{center}
\vspace*{-0.3cm}
\caption{Dependence of the partial branching fractions   
$\mbox{Br}\,(B^-\!\to\gamma\ell\bar\nu, E_\gamma> E_{\rm cut})$  
for $E_{\rm cut}=1\,$GeV (upper band) and  
$1.7\,$GeV (lower band) on $\lambda_B$. Figure from~\cite{Beneke:2011nf}.} 
\label{fig:finalBR2} 
\end{figure}

The radiative-leptonic decay is the simplest exclusive decay, and the 
$B$-meson LCDA is the only hadronic parameter relevant to this process 
at leading order in the 
heavy-quark expansion. It therefore provides a unique way to determine 
$\lambda_B$ from (future) data. The sensitivity to $\lambda_B$ can 
be seen from the leading-power expression
\begin{equation} 
F_{V,A}(E_\gamma) = \frac{Q_u m_B f_B}{2 E_\gamma\lambda_B(\mu)} \,  
R(E_\gamma,\mu) 
\end{equation}
for the $B\to\gamma$ form factors. The radiative correction factor 
$R(E_\gamma,\mu)$ and the leading 
power-suppressed effects have been investigated in~\cite{Beneke:2011nf}. 
Figure~\ref{fig:finalBR2} shows the branching fraction with a lower 
cut on the photon energy as a function of the parameter $\lambda_B$ 
and confirms the strong sensitivity. It can be shown that at sub-leading 
order in $\Lambda/m_b$ there appears a single unknown ``soft'' form factor 
similar to the soft form factor in the $B\to \pi$ transition. However, 
the difference between $F_V$ and $F_A$ is independent of this form 
factor and can be given entirely in terms of the $B$-meson decay constant 
$f_B$, at least at tree level~\cite{Beneke:2011nf}. The power-suppressed 
soft form factor limits the accuracy of the $\lambda_B$ determination. 
A numerical estimate has been obtained with the QCD sum-rule 
method~\cite{Braun:2012kp}.

\subsection{Radiative and electroweak penguin decays}

I cannot conclude this overview without a brief mentioning of 
the exclusive radiative and electroweak penguin decays $B\to K^{*} 
\gamma$ and $B\to K^{(*)}\ell^+\ell^-$, respectively. These are 
primarily driven by the two loop-induced $(\ell\ell) (s\bar b)$ 
operators in the weak effective Lagrangian with a non-negligible 
contribution from the electromagnetic dipole operator and 
four-quark operators. The hadronic matrix elements of the former 
two contributions can be parametrized by heavy-to-light form factors, 
but those of the four-quark operators are non-local. Schematically 
the transition matrix element contains two terms, 
\begin{widetext}
\begin{equation}
\langle K^* \ell\ell|H_{\rm eff}| \bar B\rangle =
\sum_i a_i(C_{7\gamma,9,10}^{(\prime)},\ldots)\,F_i^{B\to K^*}
+ 
\frac{i e^2}{q^2} \langle \ell\ell|\bar{\ell}\gamma_\mu \ell|0\rangle 
\!\int \!d^4x \,e^{i q\cdot x}\,\langle K^*| T(j_{\rm em}^\mu(x) 
H_{\rm eff}^{\rm had}(0)|B\rangle,
\label{bkllheff}
\end{equation}
\end{widetext}
\noindent
where the first comprises the electroweak penguin and electromagnetic dipole 
transitions, can be expressed in terms of form factors, and is sensitive 
to other virtual particles in the loops that generate these transitions. 
The second is non-local, dominated by QCD effects, including charmonium 
resonances in the $b\to s (c\bar c \to \ell\ell)$ transition.

Soft-collinear factorization is relevant in two ways 
in the large-recoil region 
where the kaon energy $E_{K^*}\gg \Lambda$. First, in the 
first term the QCD form factors can be expressed in terms of only 
two ``soft'' form factors $\xi_{\parallel,\perp}$ through the factorization 
theorem (\ref{BMtheorem}) for form factors. Second, and more importantly, 
the complicated second term can be calculated and expressed in terms 
of exactly the same form factors, such that the entire amplitude 
(\ref{bkllheff}) is expressed as 
\begin{equation}
\langle K^*\ell\ell |Q_i|\bar B\rangle = 
C_i \xi + \phi_B\otimes T_i\otimes \phi_{K^*} + 
{\cal O}\left(\frac{\Lambda}{m_b}\right)
\end{equation}
for every operator $Q_i$ in $H_{\rm eff}$ as shown 
in \cite{Beneke:2001at, Bosch:2001gv} for the radiative 
decays $B\to V\gamma$ and in \cite{Beneke:2001at} 
for $B\to V\ell^+\ell^-$. These papers also performed 
the $\mathcal{O}(\alpha_s)$ calculations. Note that NNLO computations 
for these decay modes are significantly harder than for charmless 
decays, since at this order three-loop diagrams must be considered 
for the vertex kernels, and two-loop diagrams for hard-spectator 
scattering. Only the simpler contributions from the dipole 
operators have been calculated up to now~\cite{Ali:2007sj}.
The results of \cite{Beneke:2001at, Bosch:2001gv} were soon generalized to 
isospin asymmetries \cite{Kagan:2001zk,Feldmann:2002iw} and the 
$B\to\rho\gamma, \rho \ell\ell$ decays \cite{Beneke:2004dp}.

\begin{figure}[t]
\vskip0.2cm
\begin{center}
\includegraphics[width=7cm]{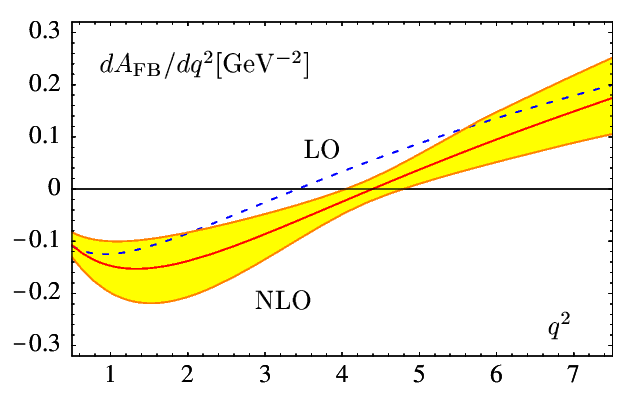}
\end{center}
\vspace*{-0.3cm}
\caption{\label{fig:dafbnorm}
Forward-backward asymmetry 
$d A_{\rm FB}(B^-\to K^{*-}\ell^+\ell^-)/dq^2$ 
at next-to-leading order (solid cent line) and leading order 
(dashed) as function of lepton invariant mass $q^2$ in GeV. 
The band reflects all theoretical uncertainties from 
parameters and scale dependence combined. Figure from~\cite{Beneke:2001at}.}
\end{figure}

As an example of the uses of factorization I show a plot from 
\cite{Beneke:2001at} for the differential forward-backward asymmetry 
already discussed above in the context of semi-inclusive hadronic final 
states. Figure~\ref{fig:dafbnorm} shows the reduction of the theoretical 
uncertainty at the location of the asymmetry zero $q_0^2$ due to 
a reduced sensitivity to the form factors \cite{Burdman:1998mk,Ali:1999mm} 
and a sizable shift of $q_0^2$ from $q_0^2=3.4^{+0.6}_{-0.5}$~GeV$^2$ at LO 
to  $q_0^2 = 4.39 {}^{+ 0.38}_{-0.35} ~{\rm GeV}^2$ at NLO. Including an 
estimate of power corrections to the form factors, 
$q_0^2=(4.2\pm 0.6)$~GeV$^2$ \cite{Beneke:2001at} remains a conservative 
estimate of the asymmetry zero location. A subsequent analysis
with updated parameters~\cite{Beneke:2004dp} gives 
\begin{eqnarray}
\label{q0result2}
&&q_0^2[K^{*0}] = 4.36^{+0.33}_{-0.31}\,\mbox{GeV}^2 
\\
&&q_0^2[K^{*+}] = 4.15\pm 0.27\,\mbox{GeV}^2 
\end{eqnarray}
for the neutral and charged $B$-meson decay separately, which replaces 
the value $q_0^2 = 4.39 {}^{+ 0.38}_{-0.35} ~{\rm GeV}^2$ above.

With the start-up of the LHCb experiment at the horizon and its large 
anticipated sample of $B\to K^{(*)} \ell\ell$ decays, the fully 
differential angular distribution including $K^* \to K \pi$ became the 
subject of investigations \cite{Egede:2008uy,Altmannshofer:2008dz}. 
From the angular coefficients one can 
construct observables with reduced sensitivity to  
$B$-meson form factors for all values of $q^2$ rather than a single 
value as is the case for the forward-backward asymmetry. This can be 
done by selecting observables that are proportional to 
only one of the products $|\xi_\parallel(q^2)|^2$, $|\xi_\perp(q^2)|^2$, 
$\xi_\perp(q^2)\xi_\parallel(q^2)$ of soft form factors, and then taking 
an appropriate ratio, such that the dependence on $\xi$ cancels out 
at leading order in $\alpha_s$. The main theoretical issue is then 
a reliable estimate of the uncertainty from $\Lambda/m_b$ corrections, 
which remains a difficult and partly speculative problem.  I refer to 
\cite{Jager:2012uw,Descotes-Genon:2014uoa} on this point and for 
many references on this very active field of LHCb physics.

\section{Summary and Outlook}

Factorization of exclusive and semi-inclusive $B$ decays has developed 
into a mature theory, making a large amount of previously intractable 
final states accessible to theoretically solid interpretations. 
The presence of energetic particles or jets implies many similarities 
with factorization methods applied in high-energy collider physics. 
A conceptually interesting point is that for exclusive decays 
the power-suppressed soft interaction
${\cal L}^{(1)}_{\xi q}=\bar q_s W_c^\dagger 
i\Slash{D}_{\perp c}\xi$~+~h.c. related to the emission (or rather 
absorption) of soft quarks $q_c \to g_{hc}+ q_s$ appears at leading 
order in the expansion of the hard scale, whereas in collider 
physics  only the familiar eikonal interaction 
$\bar{\xi} i n_- \cdot A_s \frac{\not\!n_+}{2} \xi$ describing soft 
gluon emission 
$q_c\to q_c+g_s$ is ever required at leading power. This is 
because the decaying $B$ meson provides a ``source'' of soft 
partons.

Beyond the leading power in the $\Lambda/m_b$, the available theoretical 
tools often cease to be effective. For jet-like, semi-inclusive final 
states, it appears that soft-collinear factorization can in principle 
be extended to any order as long as the invariant mass 
of the inclusive hadronic 
final state is $\mathcal{O}(m_b\Lambda)$. However, in practice already 
at the first sub-leading order in $\Lambda/m_b$ 
a large number of non-perturbative 
soft functions appears, as was discussed for the case of $\bar B\to X_u 
\ell \bar \nu$. There is presently no method that could compute these 
functions, since lattice QCD cannot be used for the matrix 
elements of non-local operators with exactly light-like field separations. 
For exclusive decays, a theory of power-suppressed effects simply does 
not exist.

Over the past ten years almost all calculations necessary to push the 
accuracy of charmless hadronic two-body decays to 
$\mathcal{O}(\alpha_s^2)$ have been performed, except for the two-loop 
QCD penguin amplitude. Once this is completed, direct CP asymmetries 
can for the first time be predicted with reliable perturbative 
uncertainties. In view of the upcoming BELLE~II experiment an update 
of theoretical predictions with NNLO uncertainty and improved hadronic 
input parameters is timely. There are still many unmeasured, but 
theoretically predicted final states. Meanwhile, the LHCb experiment 
yields data on the electroweak penguin decays 
$\bar B\to K^{(*)} \ell\ell$, the precision of which provides or will soon 
provide a 
challenge to theory. Extending the theoretical calculations 
to $\mathcal{O}(\alpha_s^2)$ is difficult here, but since the 
spectator-scattering contribution is presently known only at its 
first order, the two-loop calculation should be done.

\subsection*{Acknowledgements}

\noindent
This article summarizes work performed within and supported by the 
DFG Sonderforschungsbereich/Trans\-regio~9 
``Computational Theoretical Particle Physics''. I wish to thank many 
collaborators on this project, in particular Th.~Feldmann, T.~Huber, 
S.~J\"ager, X.~Li, J.~Rohrwild and D.~Yang. Thanks to M.~Steinhauser and 
Y.~Wang for reading the manuscript.




\bibliographystyle{elsarticle-num}

\end{document}